\documentclass[pra,aps,twocolumn,nofootinbib]{revtex4-1}

\usepackage{graphicx}
\usepackage{textcomp}
\usepackage{dcolumn}
\usepackage{bm, amssymb, amsmath, braket, dsfont,txfonts}
\usepackage{subcaption,xcolor, upgreek}
\usepackage{soul,hhline}

\begin{document}

\title{ Practical Quantum Computing:\\ The value of local computation}
\thanks{This work is licensed under a Creative Commons Attribution 4.0 International license (CC BY 4.0) \url{https://creativecommons.org/licenses/by/4.0/}}%

\author{J.R. Cruise}
\email{james.cruise@riverlane.com}
 \affiliation{Riverlane, St Andrew’s House, 59 St Andrew’s Street, Cambridge CB2 3BZ, UK}
\author{N.I. Gillespie}%
 \email{neil.gillespie@riverlane.com}
\affiliation{Riverlane, St Andrew’s House, 59 St Andrew’s Street, Cambridge CB2 3BZ, UK}%
\author{B. Reid}
\email{brendan.reid@riverlane.com}
\affiliation{Riverlane, St Andrew’s House, 59 St Andrew’s Street, Cambridge CB2 3BZ, UK}%

\date{\today}

\begin{abstract}
As we enter the era of useful quantum computers we need to better understand the limitations of classical support hardware, and develop mitigation techniques to ensure effective qubit utilisation. 
In this paper we discuss three key bottlenecks in near-term quantum computers: bandwidth restrictions arising from data transfer between central processing units (CPUs) and quantum processing units (QPUs), latency delays in the hardware for round-trip communication, and timing restrictions driven by high error rates. In each case we consider a near-term quantum algorithm to highlight the bottleneck: randomised benchmarking to showcase bandwidth limitations, adaptive noisy, intermediate scale quantum (NISQ)-era algorithms for the latency bottleneck and quantum error correction techniques to highlight the restrictions imposed by qubit error rates. In all three cases we discuss how these bottlenecks arise in the current paradigm of executing all the classical computation on the CPU, and how these can be mitigated by providing access to local classical computational resources in the QPU. 

\end{abstract}

\maketitle


\section{\label{sec:level1}Introduction}
Over the past two decades quantum computing has gained momentum as an area of theoretical research and experimental implementation. The first major milestone was reached late in 2019 with the report of so-called quantum supremacy by a research group at Google \cite{GoogleSupremacy}. The prevailing expectation is that within the next five to ten years a quantum device will exist that can perform a useful computational task that is intractable for state-of-the-art classical high performance computers (HPCs). However, this device is likely to be highly sensitive to noise and of limited scope in its applications.

\begin{figure}[t]
\centering
{
\setlength{\fboxsep}{2pt}
\setlength{\fboxrule}{2pt}
\fbox{\includegraphics[width=.8\columnwidth]{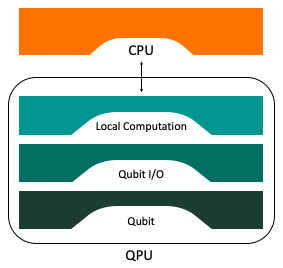}}
}
\caption{\label{fig:computer_stack} Pictorial representation of the computational stack of a quantum computer. }
\end{figure}

Current incarnations of quantum computers, like those of IBM \cite{ibm_qe} and Rigetti \cite{rigetti}, employ a `black-box' model: quantum programs are written on a central processing unit (CPU) and forwarded to the quantum hardware to be implemented blindly. This construction prevents developers having direct access to the internal components of the computer, and as such the only location for intermediate classical computation is on the user's CPU. 
Physically there are a number of intermediate platforms between the CPU and qubits that can communicate classical information with minimal disturbance, as displayed in Fig.~\ref{fig:computer_stack}. Most widespread in current hardware is the use of field programmable gate arrays (FPGAs), which sit physically close to the qubits and provide commands locally. This device's role is to control the analog hardware that directly interfaces with the qubits, for example, the lasers responsible for gate implementation. These components are unavailable to algorithm developers. We believe that providing access to these computational units will allow the maximum utilisation of noisy, intermediate scale (NISQ) devices.
NISQ-era hardware is classified by relatively short coherence times and, consequently, the need for classical hardware to supplement calculations. This hybrid quantum-classical interplay is necessary to demonstrate a computational advantage over purely classical means. However, the latency between hardware components causes qubit utilisation to suffer and a corresponding degradation of wall-clock time. If it were possible to program directly onto every component in the computational hierarchy certain latency dividends would not need to be paid. We call this `local computation': employing control protocols on local components relative to the quantum hardware. Practically, this involves identifying aspects of the desired algorithm which could be moved away from the CPU to lower levels in the stack.

In this paper we highlight three challenges in near-term hybrid quantum technology which could benefit from employing local control protocols.  These are: a \emph{bandwidth} bottleneck due to the limited communication capacity between the CPU and quantum processing unit (QPU);
a \emph{latency} bottleneck due to the round trip delay between the CPU and QPU;
and the \emph{qubit error rate} due to the short coherence time of qubits.
We propose that one possible method for overcoming these challenges is to utilise latent computational power. By identifying simple calculations that need to be performed regularly (such as a parameter update or logical \texttt{while} loop) these can be assigned to hardware components physically close to the qubits: the remaining difficulty is that this task is non-trivial and beyond the skill set of a typical programmer. While we will not propose a concrete pathway to solving these issues, we aim to convince the reader that developing a more hardware-conscious methodology is a priority for the immediate future of quantum computation.

\subsection{Hardware Stack}\label{sec::hardware}
Whilst quantum computers utilise a variety of different technologies and implementations, they follow a similar structural design. This begins with the actual qubits at the bottom of a stack, continues through the hardware used to physically interact with the qubits (qubit I/O), onto the computational units used to control the physical hardware, and on top a classical computer that we refer to as the CPU (see Fig. \ref{fig:computer_stack}). This stack can be seen when visiting hardware laboratories working on a diverse range of qubit technologies, including trapped ions, superconducting qubits, and semiconductor qubits. 

We will focus on the intermediate hardware between the QPU and CPU. As mentioned the most common hardware used to interface between the CPU and QPU is FPGAs. However, there are alternative programmable platforms that may become more popular in the future. While the important feature of this hardware is its programmability, equally important is its physical connection to both the CPU and QPU. Currently there are a number of technologies that fill this role within the hardware stack, including Ethernet, universal serial bus (USB) and peripheral component interconnect express (PCIe). Each technology has benefits and limitations both in regards to capacity and ease of use. For example, Google has demonstrated the use of Ethernet for this connection within a cloud computing setting in their recent work on optimising variational quantum eigensolver (VQE) implementations \cite{sung2020exploration}. In comparison many laboratories are using the PCIe lanes in commercially available motherboards.

The two metrics we will be concerned with in this paper are the communication latency (the delay between sending and receiving a message across the link) and the available bandwidth (the rate at which data can be sent between the two devices). For communication latency, we will report results based on a timing of $100\upmu$s for one-way communication -- we consider this a realistically achievable value without substantial engineering overhead, though we also include plots which highlight the effect of latency so other values can be easily considered. For bandwidth, we make no assumptions but will report requirements for the communication channel to prevent a degradation in performance. This will more clearly allow practitioners to understand where the bottlenecks will form within their system.   



In terms of physical qubits there are a number of competing technologies; here we will focus on timings for superconducting qubits and trapped ions. The reason for this is that superconducting qubits demonstrate relatively fast gates times but short-lived coherence whereas, in comparison, ion traps have substantially longer coherence accompanied with very slow gate implementation. The majority of other qubit technologies lie between these two extremes. As we will focus on quantifying overall performance we will consider only gate times rather than the details of qubit implementation. Throughout this work we consider gate and measurement times to be 10$\upmu$s and $750\upmu$s respectively for trapped ion qubits~\cite{tion}, and 120ns and 120ns for superconducting qubits ~\cite{superconducting}. Note that these are relatively long and we expect these times to reduce as further research is conducted. 



\subsection{Computational Challenges}

 \begin{figure*}[t]
\centering
{
\setlength{\fboxsep}{5pt}
\setlength{\fboxrule}{1pt}
\includegraphics[width=.9\textwidth]{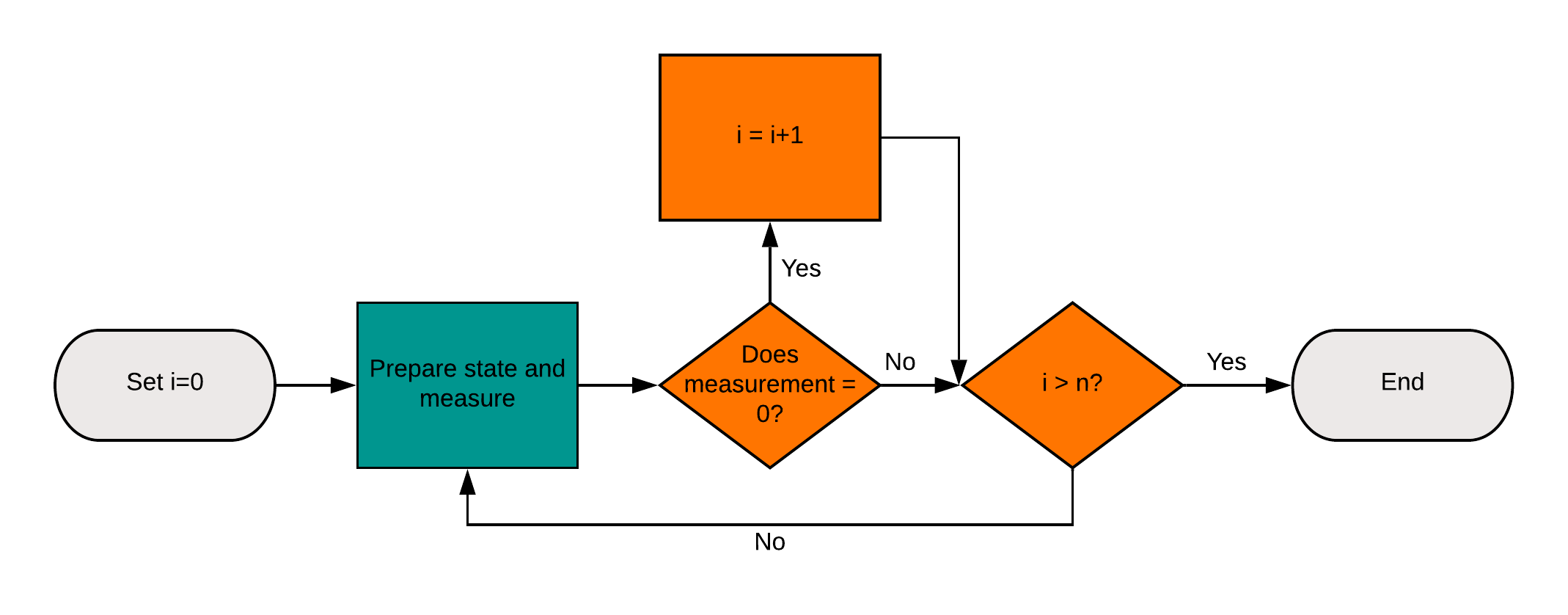}
}
\caption{\label{fig::while} Simple while loop which stops only when $n$ measurements of 0 have been completed. Under the black-box model, the green boxes will be completed on the QPU and the orange on the CPU.}
\end{figure*}

To highlight the issues of carrying out all classical computation on the CPU we consider a toy example -- a while loop. The loop runs over a circuit execution and measurement, counting the number of times the value 0 is returned and stopping when we have reached a certain count (see Fig. \ref{fig::while}).  This example might seem overly simple, but quantum expectation estimation to a high confidence will be similar to this setup.

In this toy example a message will be have to be passed to the CPU between each circuit execution to decide if another run is required or not. As mentioned above this will add a delay of approximately $200 \upmu$s for each circuit evaluation. The time required for a measurement to be generated will be $O(1)\upmu$s, resulting in the qubits sitting idle for $99\%$ of the computation time waiting for the CPU to issue commands. Even for trapped ion qubits, where the circuit execution time is likely to be on the order of $800 \upmu$s, this will still lead to the qubits sitting idle for approximately 20\% of time. If the update and checking process were completed locally on an FPGA or similar device, qubit utilisation would be increased substantially.   

\vspace{12pt}
Randomised benchmarking \cite{knillrb, originalrb, flammia1, ProctorRB} is a method to provide summary statistics about inherent noise levels in quantum devices, for example, average gate fidelity, which is key for understanding device performance. Zero noise extrapolation \cite{OZNE,benjaminzne,kandala,digital} is an error mitigation scheme. Both algorithms need to run a large number of random circuits and record the results. If circuit generation occurs on the CPU, rather than locally relative to the qubits, a substantial amount of data must be transferred to prevent low qubit utilisation from lack of circuits to execute. We discuss this in greater detail in Section \ref{sect:rb}.

There has been a growing interest in adaptive refinements to quantum algorithms in order to better utilise quantum devices. In these procedures the circuit to be implemented and relevant statistics are updated between circuit executions as in the while loop example. For instance, accelerated VQE (AVQE) \cite{Daochen} introduces an adaptive quantum subroutine to outperform VQE in energy minimisation. Section \ref{sec:AVQE} goes into detail of how this subroutine operates. We note that AVQE is not unique in modifying VQE for improved performance using adaptive routines; work in the same spirit was recently released by Zapata Computing \cite{zapataELF}.

Implementing error correcting strategies in quantum devices, both in the near term \cite{holmes} and when lower physical error rates have been achieved \cite{delfosse}, is critical. In order to avoid an exponential growth of noise, errors must be dealt with frequently. Any correction protocol needs to be fast enough
to keep up with the error rate otherwise a data bottleneck is created, which can result in the loss of any quantum advantage \cite{holmes}. See Section \ref{sect:qec} for further details on quantum error correction.

Finally, in Section \ref{sec:conclude} we provide some concluding remarks and comments on providing developers with access to local computation.


\begin{figure*}[t]
\centering
{
\setlength{\fboxsep}{5pt}
\setlength{\fboxrule}{1pt}
\includegraphics[width=.9\textwidth]{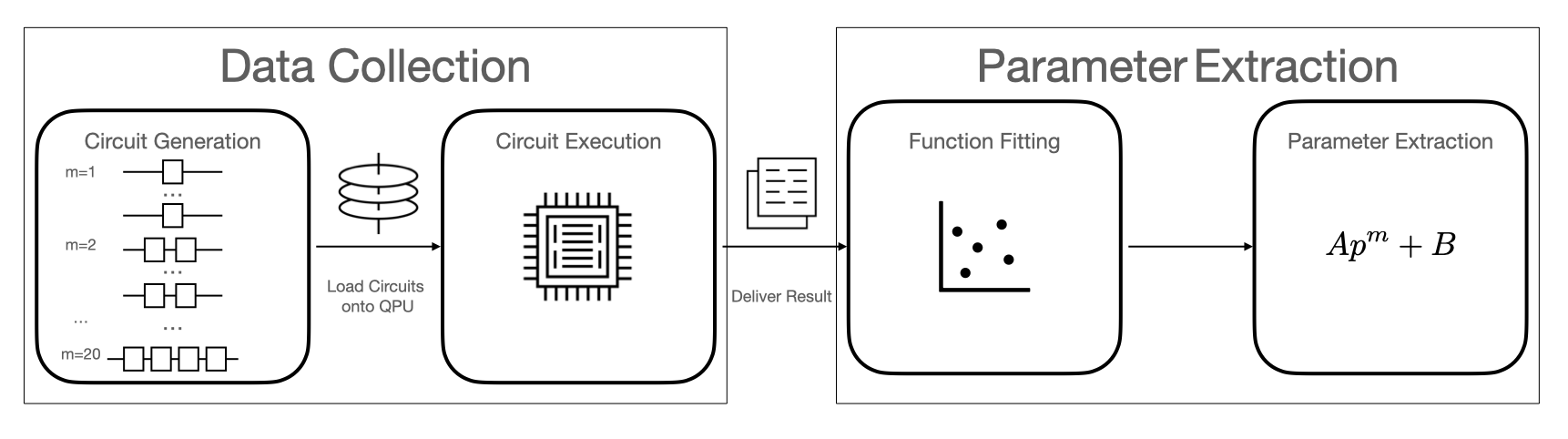}
}
\caption{\label{fig::alg_diag}Diagrammatic representation of algorithms (Randomised Benchmarking or Zero Noise Extrapolation) }
\end{figure*}

\section{Randomised benchmarking and Zero Noise extrapolation}\label{sect:rb}

In the NISQ-era of quantum computing one of the key challenges will come from noise inherent in the quantum computer. To maximise the utility of these computers we will need to both quantify the level of noise within the device and develop noise mitigation strategies. Both of these are key tasks but they are difficult to achieve in a scalable manner, often growing exponentially in computational complexity  with the number of qubits \cite{tomography}. In this section we will discuss two protocols which have gained substantial traction within the quantum computing community as their computational complexity scales more favourably with the number of qubits: randomised benchmarking \cite{originalrb, originalrb2,knillrb, ProctorRB} for noise quantification and zero noise extrapolation\cite{OZNE,benjaminzne,kandala,digital} for noise mitigation. In both cases we highlight how the algorithms naturally lead to a bandwidth bottleneck which can be removed through the use of local computation. 

Randomised benchmarking was initially proposed as an alternative to quantum tomography \cite{tomography} as a methodology to quantify the quality of the computational capabilities of quantum computers. Unlike quantum tomography, randomised benchmarking does not provide a full specification of the action of various gates but instead obtains estimates of summary statistics, for example, average gate fidelity.  In exchange for obtaining lower quality information, we obtain a protocol with substantially improved scaling as we increase the number of qubits and that is robust to state preparation and measurement (SPAM) errors. Randomised benchmarking has gained traction within the experimental quantum computing community and has been the focus of a concerted research effort. This has included exploring adaptive algorithmic improvements \cite{granade,harper,flammia1}, different gates sets \cite{othergate1,othergate2} and error models \cite{carignan, wallman}, generalisations to capture further details of quantum computing performance \cite{cycle,hashagen}, and further theoretical understanding of algorithmic performance \cite{flammia1, harper, ProctorRB}.

 Zero noise extrapolation \cite{OZNE,digital} is a class of algorithms and techniques to mitigate noise in near term quantum computers. For details of alternative strategies see \cite{cai,mcardle,benjamin2}.  In zero noise extrapolation we take the counter-intuitive step of actively increasing the noise in the calculation and recording the outcome over a large range of noise levels.  This data is then used to extrapolate back to the noiseless situation which is, in general, unobtainable from direct calculation. This works under the assumption that the calculation degrades smoothly and constantly as the noise level increases, so the degradation at higher noise levels can shed light on the performance at low noise levels.  This was initially proposed by \cite{OZNE} using Richardson extrapolation. Later work has explored alternative methods for increasing the noise level \cite{OZNE,he,kandala}, alternative extrapolation methods \cite{benjaminzne, benjamin2,digital} and the development of an adaptive framework \cite{digital}.

A key step in both algorithms is data collection, see Sect. \ref{sec::alg}, which involves running a large number of different circuits on the qubits in question and recording the results. If the circuit generation occurs on a CPU, rather than local to the qubits, a substantial  amount of data must be transferred between the two devices  and at speed to prevent low qubit utilisation,  see Sect. \ref{sec::bbneck}.  This transfer causes a substantial bandwidth bottleneck between the CPU and the QPU. This bottleneck can be removed by moving the circuit generation process to the QPU's local computation hardware, see Sect. \ref{sec::localgen}, therefore substantially reducing the communication overhead between the CPU and QPU and increasing qubit utilisation.

Furthermore, recent developments have introduced adaptive alternatives for both randomised benchmarking \cite{flammia1, granade} and zero noise extrapolation \cite{digital}. Adaptive algorithms in general suffer from poor performance due to high latency between the CPU and QPU. In this section we will briefly touch on the use of local computation to mitigate this poor performance, Sect.~\ref{sec::rblocalupdate} , but this will be dealt more fully  in the section discussing the Accelerated Variational Quantum Eigensolver, Sect.~\ref{sec:AVQE}.

\subsection{Algorithm Description}\label{sec::alg}
 Both algorithms in their simplest form follow two steps: 
 \begin{enumerate}
 	\item {\bf Data collection:} Independent identical realizations of random circuits, indexed by a parameter $m$, are applied to a prepared input state and the measurement outcome recorded. Data collection is carried out over a range of values of $m$. 
 	\item {\bf Parameter estimation:} A function of the control parameter $m$ is fitted to the estimated proportion of measurements returning a given state from the data collection. The parameters of the fitted function provide estimates for the key statistics of interest. For randomised benchmarking we estimate the average gate error rate and for zero noise extrapolation we estimate the true value of the function. 
 \end{enumerate} 

 See Fig.~\ref{fig::alg_diag} for  a diagrammatic representation of the process: the data collection including random circuit generation and evaluation on qubits, then parameter estimation including function fitting and parameter extraction.  Initially, we will consider these two steps run in a sequential fashion where the parameter estimation step does not influence the data collection, i.e. the number of samples at each $m$ level is predetermined. In Sect.~\ref{sec::adapt} we will discuss recent developments of adaptive parameter estimation schemes.

\subsubsection{Randomised benchmarking}\label{sec::RBAlg}
Randomised benchmarking is used to estimate the average gate fidelity for a given set of qubits through the application of a sequence of random gates designed to implement the identity operation if evaluated on an error free set of qubits.  

\begin{figure}[t]
\centering
{
\setlength{\fboxsep}{5pt}
\setlength{\fboxrule}{1pt}
\includegraphics[width=.8\columnwidth]{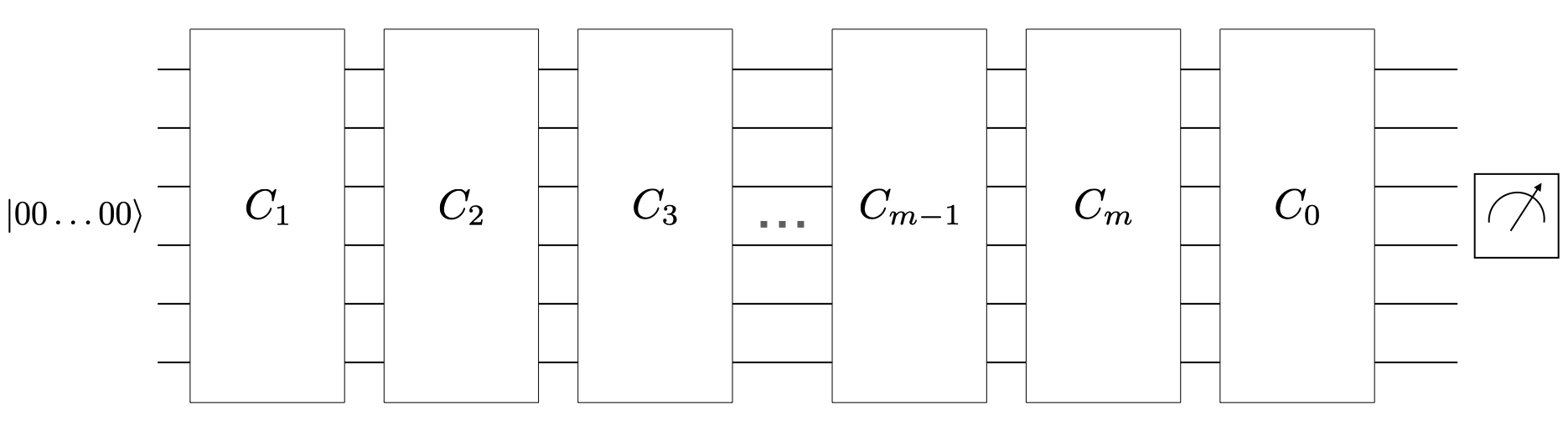}
}
\caption{\label{fig::rbcircuit}Example Circuit for Randomised benchmarking, $C_1, C_2, \ldots, C_{m-1}, C_m$ are random Clifford gates and $C_0$ is the cumulative inverse of $C_1 C_2 \cdots C_{m-1} C_m$.}
\end{figure}

{\bf Data Collection:} For a given circuit depth $m$ we generate a sequence of $m$ random Clifford gates, $C_1, C_2, \ldots C_m$  on the $n$ qubits we wish to benchmark. The random Clifford gates are sampled independently from the set of all Clifford gates on $n$ qubits, see \cite{rclifford1,rclifford2} for an example scheme to do this.  The inverse gate of the composition of these $m$ gates  is then calculated: $C_0= (C_1 C_2 \ldots C_m)^{-1}$. The circuit to be evaluated is then the sequential application of these $m+1$ gates to an input state and then measurement in the computational basis, with the goal of comparing the initial and final state. In most circumstances the input state will be the all zero state $\ket{0^{\otimes n}}\equiv\ket{00\!\ldots\!0}$.  See Fig.~\ref{fig::rbcircuit} for a diagrammatic example of such a circuit. This process is repeated for each $m$ value a large number of times, producing a number of independent samples, and then across a large range of $m$ values. 

 Initial descriptions of randomised benchmarking were based on samples from the full unitary group rather than restricting to only the Clifford group \cite{2design, magesan}, as the aim was to estimate the average gate fidelity over the entire gate set rather than just the Clifford subset. The discovery that the Clifford group forms a 2-design for the unitary group \cite{2design} means that restricting to the Clifford gate set  still provides an estimate of the average fidelity over the whole unitary group. As such, we will discuss randomised benchmarking protocols that consider the Clifford group only.
 
 {\bf Parameter estimation:}  Now we want to use the acquired data to estimate the average gate error rate and by extension the average gate fidelity, a key indicator of qubit performance.  We will assume that gate errors are independent and homogeneous in time, which is often referred to as the $0^{\mathrm{th}}$ order model. More complicated models have been developed, for details see \cite{magesan}. Using this assumption, we want to fit a model to the estimated survival probabilities, $\bar{p}_m$, the proportion of circuit runs returning the originally prepared state after $m$ random Clifford gates and the associated cumulative inverse. 

\begin{figure}[t]
\centering
{
\setlength{\fboxsep}{5pt}
\setlength{\fboxrule}{1pt}
\includegraphics[width=.8\columnwidth]{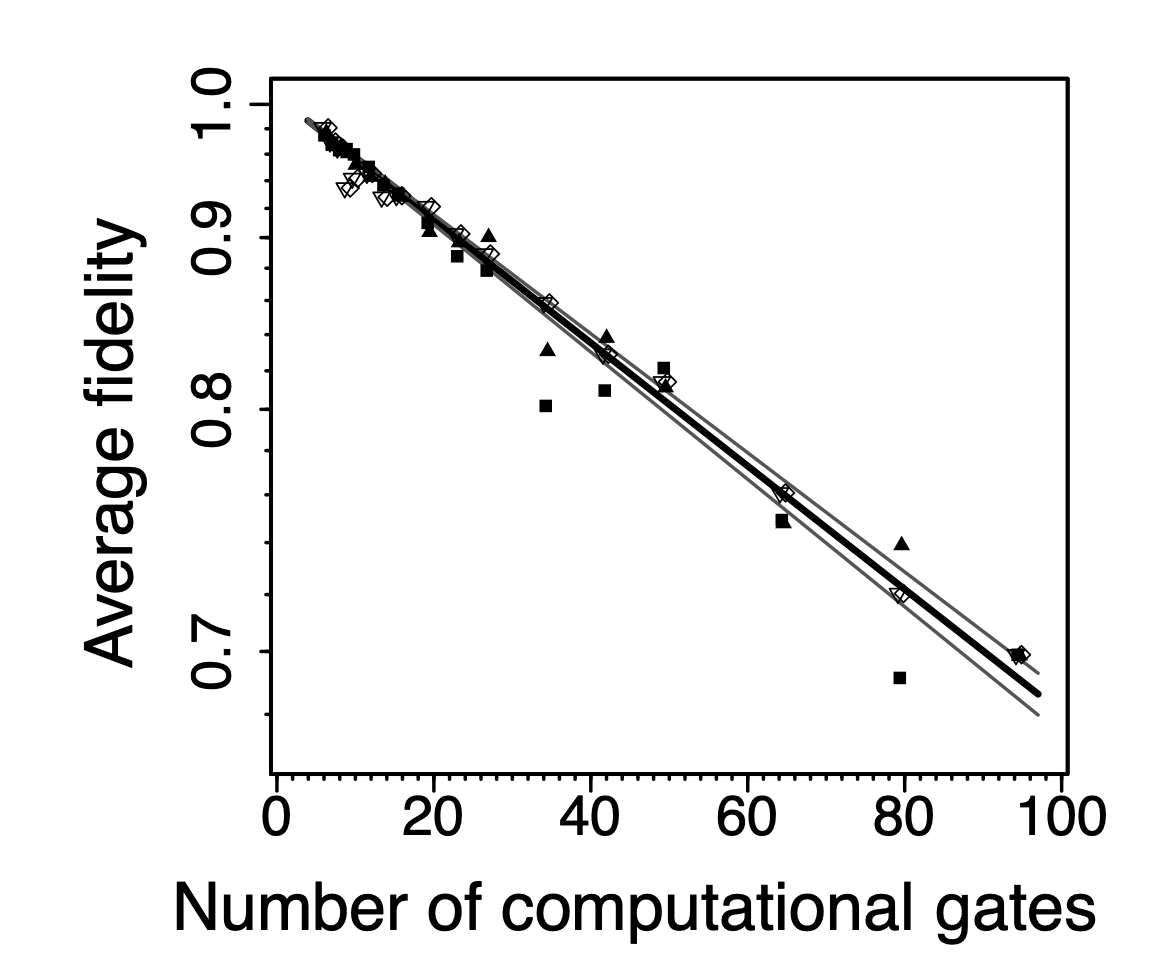}}
\caption{\label{fig::rbexample} Example of Randomised benchmarking data and function fitting taken from \cite{knillrb} }
\end{figure}

 Under the time and gate independence assumption it can be shown that the application of $m$ independent random Clifford gates and their cumulative inverse is equivalent to applying $m$ independent copies of the depolarizing channel to the initial state with a depolarizing probability $p$ ---  the average gate error rate.  Therefore the probability of recovering the original state is $A+Bp^m$, a geometric probability, where $A$ and $B$ allow for state preparation and measurement errors and $p$ is the average gate error rate. Hence we fit this model to the estimated survival probabilities, normally by fitting a linear equation to $\log(\bar{p}_m)$ and using the estimated slope to calculate the average error probability.  An example of this for real data can be seen in Fig.~\ref{fig::rbexample} from \cite{knillrb}.
 
\subsubsection{Zero noise extrapolation}\label{sec::ZNAlg}
In zero noise extrapolation, we carry out the calculation of interest across a wide range of noise levels by artificially increasing the noise and using the estimated values to extrapolate back to the case with zero noise.   

\begin{figure}[ht]
\centering
{
\setlength{\fboxsep}{5pt}
\setlength{\fboxrule}{1pt}
\includegraphics[width=.8\columnwidth]{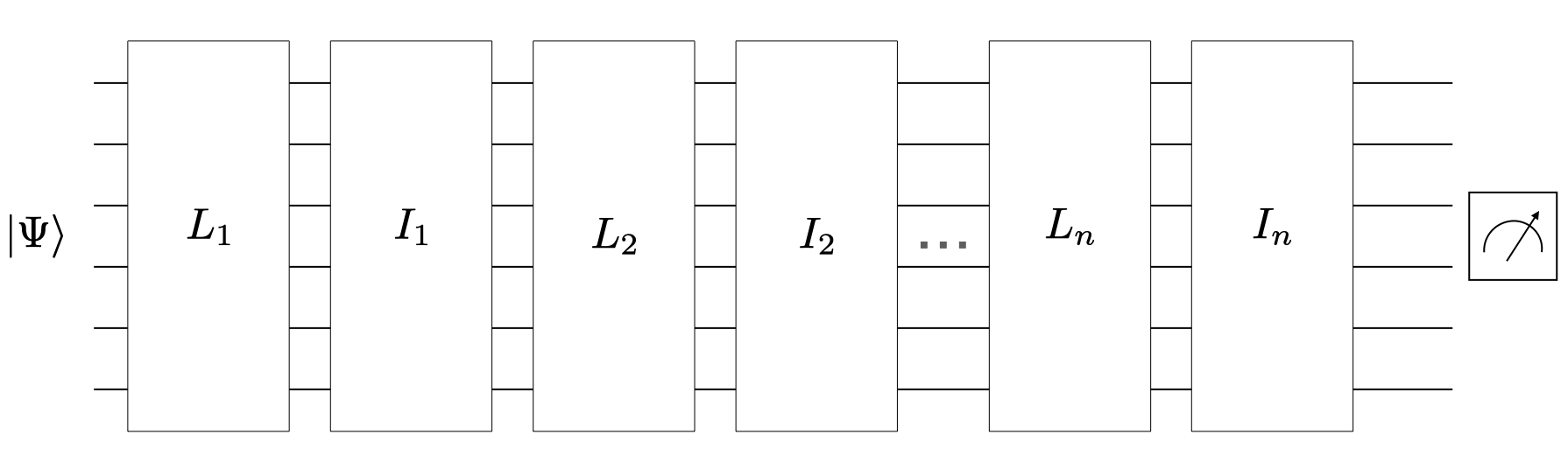}
}
\caption{\label{fig::inter}Example Unitary Folding Circuit, $L_1,L_2,\ldots,L_n$ are layers of gates implementing the original circuit and $I_1,I_2, \ldots, I_n$ are identity blocks such that $I_i = U_i U_i^\dagger$ where $U_i$ are randomly sampled unitary gates.  }
\end{figure}

{\bf Data Collection:} Given a circuit to estimate an expectation value $E$, we want to implement this circuit at a wide range of noise levels. There are number of ways to artificially increase the  noise level including unitary folding and parameter noise scaling \cite{digital} . For unitary folding, see Fig.~\ref{fig::inter}, we consider the circuit to be a series of $n$ unitary layers
 $L_1, L_2, \ldots L_n$, and between each layer there are identity operations  consisting of a random unitary gate and its inverse, $I_i=U_i U_i^\dagger$. The noise level is increased through the addition of extra identity blocks between layers.  In parameter noise scaling, for every gate  parameterized by $\theta$, we now implement the gate using a perturbed  value, $\theta' = \theta +X$ where $X$ is randomly generated and $X\sim\mathcal{N}(\mu = 0,~\sigma^2)$.  Here the level of artificial noise is controlled by the variance of the perturbation. 
 
In both scenarios we carry out the expectation estimation across a range of noise \emph{levels}, treating the inherent noise  of the quantum computer as noise \emph{level one}. For each noise level we execute a large number of  independent identical samples of the circuit, i.e. for unitary folding this would be with newly sampled independent unitary gates  or for parameter noise scaling resampling the perturbation for each parameter $\theta$. The results of the samples are then used to produce empirical estimates of the expectation values, $\bar{E}(\lambda)$, where $\lambda$ is the noise level, which are then used in the parameter estimation step.

{\bf Parameter estimation:} Parameter estimation in zero noise extrapolation aims to fit a function to the expectation estimates, $\bar{E}(\lambda)$,  across a range of noise levels $\lambda>1$.  This fitted function can then be used to extrapolate for $\lambda=0$ and therefore $E(\lambda = 0)$, the value of the expectation without noise. The hope is that the true expectation value will be  smooth as we vary the noise parameter and there will not be a substantial change in behaviour for $\lambda<1$.

In general, there is no a priori class of functions to fit to the data as it is unknown how the noise will effect the expectation value. A number of general schemes have been used, including Richardson extrapolation \cite{OZNE},  polynomial, exponential and  linear regression \cite{benjamin2,digital}.

\subsubsection{Adaptive modifications}\label{sec::adapt}

In the discussion of the algorithms, we split them into two distinct steps: data collection and parameter estimation. We complete the data collection process before carrying out the parameter estimations. This means that we use a pre-defined data collection strategy: specifying the parameter range to consider and the number of samples to be collected for each value in that range.

Unlike most statistical settings, there is no need for this artificial separation; instead, the data collection strategy can be updated in an adaptive manner to maximise the obtained information given the current parameter estimates. In this case, we would determine at each circuit execution what value of parameter to use for sampling to maximise the acquired data or minimise the total number of computational resources required to obtain a given accuracy. This leads us to develop the sampling schedule in an online fashion, as we collect the data  and carry out an update step between shots. For example, such adaptive algorithms can be seen in \cite{granade, flammia1} for randomised benchmarking and \cite{digital} for zero noise extrapolation. 

\subsubsection{Repeated circuit evaluations}\label{sec::repeat}

\begin{figure}[t]
\centering
{
\setlength{\fboxsep}{5pt}
\setlength{\fboxrule}{1pt}
\includegraphics[width=.8\columnwidth]{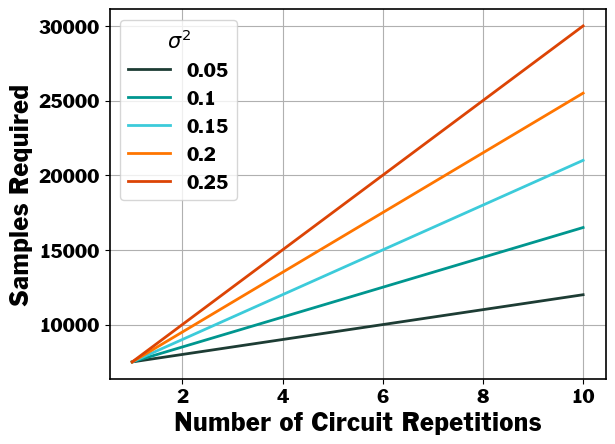}}
\caption{\label{fig::repeat}Number of samples required to obtain an accuracy of 0.01 for $\mu=0.5$, the true value, across a range of $\sigma^2$ values, the between circuit noise. }
\end{figure}

In the described data collection steps for both randomised benchmarking, Sect.~\ref{sec::RBAlg}, and zero noise extrapolation, Sect.~\ref{sec::ZNAlg}, for each circuit execution a new random circuit was generated. However, the noise in the estimation comes from two sources: sampling from random circuits and the inherent shot noise.  Therefore, an alternative strategy is to execute each random circuit a number of times rather than just once, hopefully lowering the number of random circuits that need to be created. Unfortunately, this increases the total samples (hence number of circuit evaluations) required to obtain the same accuracy. See Fig.~\ref{fig::repeat} for number of samples required to obtain the same accuracy in the toy example described below.  

Consider the following toy model, let $X\sim \mathrm{Bernoulli}(P)$, where $P$ is a random variable such that $\mathbb{E}(P) = \mu$ and $Var(P)= \sigma^2$. Here, $\mathbb{E}(X)= \mu$, is the expectation we want to estimate, and $P$ represents the noise between circuits.  We acquire samples of $X$ and we want to estimate the true value of $\mu$. We now consider two schemes. In the first scheme, we produce $k$ independent samples of $P$, $\{P_i\}_{i=1\ldots k}$, and then for each sample $P_i$ we produce $l$ independent samples of $X|P_i$, $\{X_{i,j}\}_{j=1,\ldots l}$. In the second scheme, we instead produce $n=l\times k$ independent samples of $X$, the same number of samples as in the previous scheme, but for each sample of $X$ we use a new sample of $P$. It should be noted that the first scheme  reuses the random circuits, while the second scheme produces a new random circuit for each shot. In both cases we estimate $\mu$ by the empirical mean of samples of $X$, $\bar{\mu} = \frac{1}{l k} \sum X_i $. 

To compare the two schemes we are interested in the variance of the estimator. For the first sampling scheme we have that
$$\mathrm{Var}(\bar{\mu}_1) = \frac{1}{l k}\left[\mu-\left(\mu^2+ \sigma^2\right)\right] + \frac{\sigma^2}{k}.$$
For the second sampling scheme we have
$$\mathrm{Var}(\bar{\mu}_2) = \frac{1}{l k}\left[\mu-\left(\mu^2+ \sigma^2\right)\right] + \frac{\sigma^2}{l  k}  = \frac{\mu -\mu^2}{n}.$$
Therefore, we have $\mathrm{Var}(\bar{\mu}_1) \geq \mathrm{Var}(\bar{\mu}_2)$ with equality only if $\sigma$ = 0 or $l=1$. We achieve a smaller variance if a new random circuit is obtained for each shot rather than reusing circuits. To further highlight this, in Fig.~\ref{fig::repeat} we plot the total number of samples required to obtain a given accuracy as a function of $m$, the number of times we reuse the circuit.      

\subsection{CPU to QPU bottlenecks}

Many implementations of these algorithms utilise a CPU and QPU model for their execution i.e. where all the classical computation is carried out on the CPU and the QPU is used purely to execute the prepared circuits. As discussed in the introduction this model fails to take full advantage of the computational resources available  much closer to the qubits, often located on FPGAs tasked with hardware control. Here we will highlight two key bottlenecks in the performance of both randomised benchmarking and zero noise extrapolation which come about because of this design decision. Then, in the next section, Sect~\ref{sec::rblocal}, we explain how utilisation of the local computation can improve performance through the removal of these bottlenecks. 

\begin{figure*}[t]
\centering
{
\setlength{\fboxsep}{5pt}
\setlength{\fboxrule}{1pt}
\includegraphics[width=.9\textwidth]{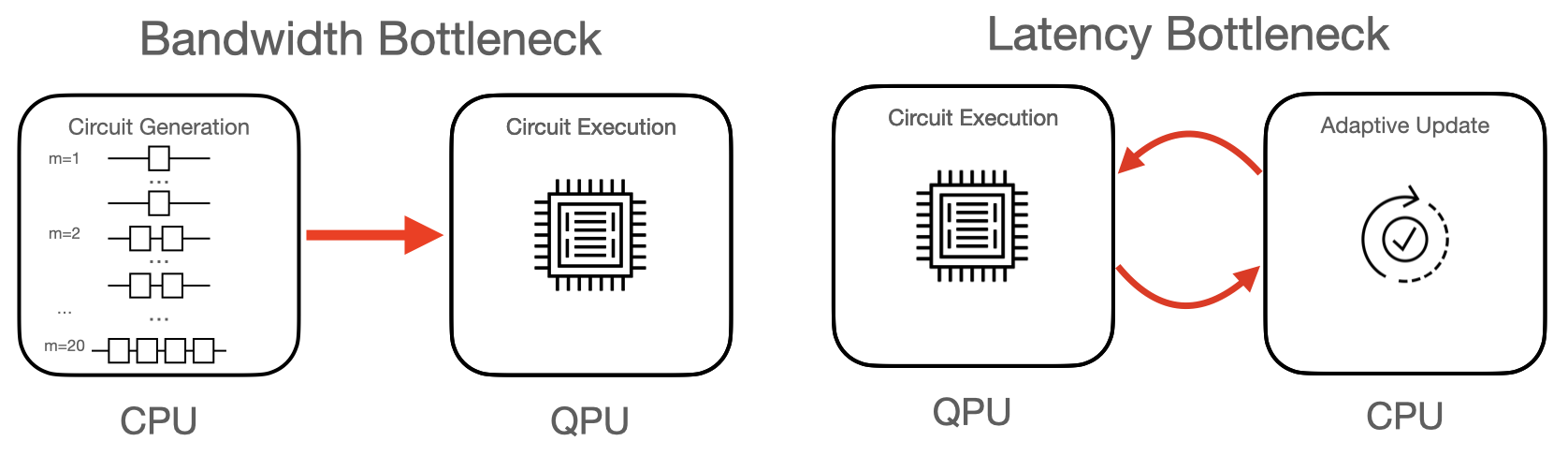}
}
\caption{\label{fig::bb} Bottlenecks considered in this section. The orange arrows represent the bottlenecks, Bandwidth bottleneck between circuit generation (left) and execution and Latency bottleneck between circuit execution and adaptive update (right).}
\end{figure*}

The first of these bottlenecks is due to the limited bandwidth available between the CPU and QPU control hardware, see Fig.~\ref{fig::bb}. This comes from the need to provide a constant stream of different circuits to be executed on the QPU within the data collection phases of the algorithms. The second of these is due to the high latency involved in the round trip of carrying out an update within an adaptive algorithm, i.e. the circuit to be run next depends on the outcome of the current circuit. The relatively high communication lag will often be an order of magnitude longer than the circuit execution and computational update step times leading to a substantial degradation in performance, see Fig.~\ref{fig::bb}.  
 
\subsubsection{Bandwidth Bottleneck} \label{sec::bbneck}

In both randomised benchmarking and zero noise extrapolation one of the key challenges is to execute a large number of different circuits. This contrasts with VQE, \cite{PeruzzoVQE}, which executes the same circuit a number of times. As highlighted in Sect~\ref{sec::repeat}, to minimise the total number of circuit executions and therefore total run time of the algorithms, each generated circuit can only be run once on the qubits. To understand the bandwidth requirement of delivering such a stream of circuits from the CPU to QPU, such that this does not create a bottleneck, we consider a slightly simpler case to analyse. Namely, the problem of providing a constant stream of gates to all qubits within the system.

\begin{figure}[t]
\centering
{
\setlength{\fboxsep}{5pt}
\setlength{\fboxrule}{1pt}
\fbox{\includegraphics[width=.8\columnwidth]{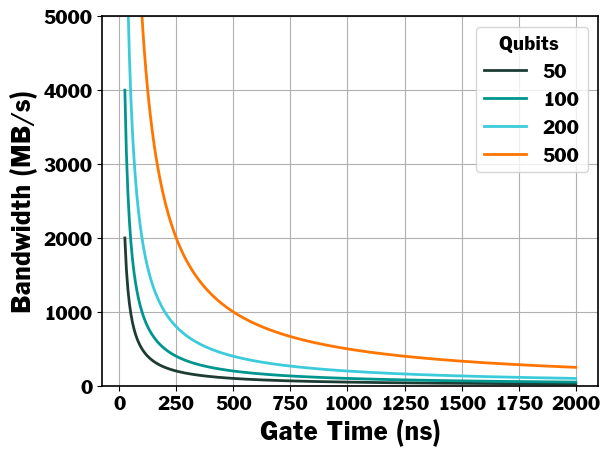}}
}
\caption{\label{fig::bandwidth}Bandwidth requirements for a constant gate stream at 50\% utilisation.}
\end{figure}

We  assume a data requirement  of 2 bytes of information to specify a gate per qubit. For a 2-qubit gate this means 2 bytes to specify the qubits the gate acts upon and 2 bytes to specify the action gate, a total of 4 bytes. For the 1-qubit gates this is a byte to specify the qubit the gate acts upon and a byte to specify the gate action. In Fig.~\ref{fig::bandwidth} we plot the bandwidth requirement to deliver such a gate stream against average gate time for a range of device sizes assuming a 50\% qubit utilisation, meaning that a qubit will only be involved in 50\% of the layers of gate execution. From this we can see that for a superconducting system with an average gate time of $120ns$ with 150 qubits this will require a bandwidth of at least 1.2GB/s. Note that this is under a relatively low qubit utilisation assumption and a gate time which is only likely to improve substantially in the coming years, both of which will further increase this already high communication requirement. Furthermore, this calculation ignores the communications overheads required to transfer data between the CPU and the QPU control hardware; these will further inflate the communication requirements. Maintaining data transfer rates of this level is a substantial engineering challenge and in many cases will be unattainable. Failure to obtain these data rates will lead to under-utilisation of qubits and increased running time for the various algorithms.

\subsubsection{Latency Bottleneck}

As describe in Sect. \ref{sec::adapt} there is a move to improve performance, accuracy and run time, through the use of adaptive algorithms. In adaptive algorithms  the circuit specification, either depth for randomised benchmarking or noise level for zero noise extrapolation, is updated at every shot. This requires a round-trip to the CPU between each shot within this computational model, see Fig.~\ref{fig::bb}. The associated communication latency of such a round trip is often on the order of $200 \upmu s$. See Sect. \ref{sec::hardware} for a discussion of CPU and QPU communications, which are 40 times longer than the $5 \upmu s$ required to execute a circuit and update the circuit specification for superconducting qubits. Therefore the communication latency will lead to a substantial degradation in algorithmic performance and low qubit utilisation. See Sect. \ref{sec:AVQE} and specifically Sect. \ref{sec::cost_subroutnine} for a more detailed discussion of adaptive algorithms and the associated bottleneck. 

\subsection{Local computation mitigation}\label{sec::rblocal}
Having discussed the bottlenecks, bandwidth and latency, which arise from the traditional model of carrying out all traditional computation on the CPU rather than locally to the qubits, we now highlight how the use of computation local to the qubits can mitigate these and remove the bottleneck.  

\subsubsection{Local circuit generation} \label{sec::localgen}
 As described in Sect.~\ref{sec::alg} for both randomised benchmarking and zero noise extrapolation, the random circuits are sampled from a predefined template. Using these templates, the creation of the next circuit to be run at each time step is a relatively low weight computational task. Rather than completing this task using the CPU, this can instead be completed, with a low overhead, on the simple computational units making up the available local computation.  For example, within near term quantum computers, the FPGA used for hardware control can be programmed to carry out this task. The use of the available local computation in this fashion completely removes the bandwidth bottleneck. As the local computation directly interfaces with the qubit hardware, the only data transfer is the circuit results and, depending the on set up, a stream of random numbers.  
 
 \subsubsection{Local update}\label{sec::rblocalupdate}
 
 As will be discussed in detail for the AVQE in the next section, for adaptive algorithms in general we can carry out the required update calculation and decision on the local computation hardware rather than the CPU. There is no need to communicate back to the CPU and wait for a response between circuit evaluations. This removes the communication delay completely, hence reducing the length of time required to complete each iteration of the algorithm and therefore the total run time. 
 
 It is worth noting that the computational resources available on local computation hardware are often limited. For example, in near term devices this will generally be the number of FPGAs. Often this means that developers will implement approximations of the full update process that would be carried out on the CPU or develop lighter weight algorithms that give close to optimal performance. This may lead to a slight degradation in algorithmic performance but, in general, this will be easily compensated for by the improved performance due to removal of the latency bottleneck.


\section{Accelerated Variational Quantum Eigensolver}\label{sec:AVQE}
\subsection{Introduction}
Variational algorithms provide a promising pathway for investigating quantum effects on near-term (NISQ) quantum hardware. The hybrid quantum-classical nature of these algorithms means that required coherence times on a QPU are minimised and a classical component can perform the bulk of the necessary work. The VQE is the flag-bearer for this class of algorithms and has found several applications in quantum chemistry \cite{OMalleyPRX, KandalaNature,ParrishPRL}, which is widely expected to be one of the first use-cases for quantum computing machines. While QPU time is minimised by using only short quantum circuits, there is a penalty in the number of times those circuits actually need to be implemented. Specifically, the circuit being implemented has depth $d= O(1)$ but to achieve a required precision $p$ the total number of circuit executions (equivalently, measurements) is $O(p^{-2})$ \cite{PeruzzoVQE}. VQE was developed for use in estimating ground state energies of quantum Hamiltonians as an alternative to Kitaev's quantum phase estimation (QPE) algorithm. While Kitaev's algorithm has advantages in that the number of measurements required is $O(1)$, the required circuit depth is beyond the scope of NISQ devices --- $O(p^{-1})$ for precision $p$ \cite{AG-QPE,LloydQPE}. 

\begin{figure}[t]
\hspace{-20pt}\includegraphics[width=.95\columnwidth]{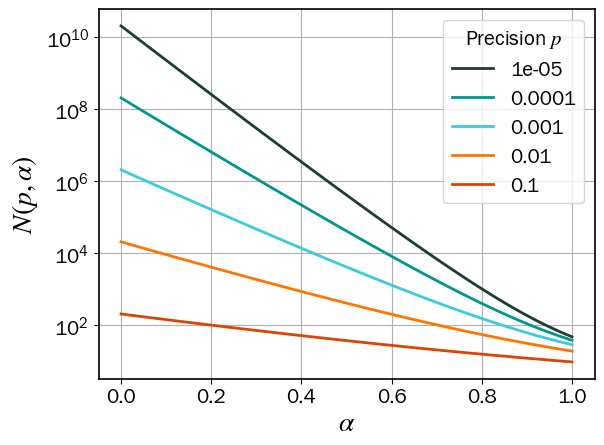}
\caption{\label{fig:measurements}Measurements for the AVQE algorithm $N(p, \alpha)$ plotted against $\alpha$ for various values of the precision $p$. }
\end{figure}
These two algorithms provide recourse in two different hardware regimes; the gap between $O(1)$ and $O(p^{-1})$ coherent circuit depth is vast and the technological bridge between them will be made incrementally. 
This realisation has spurred development into algorithms that allow circuit depth as a (directly or indirectly) controllable parameter \cite{bqpe, zapataELF}. In contrast to running VQE (whose circuit depth is fixed) until the requisite hardware is available to run QPE, the ability to vary the circuit depth your algorithm requires will allow more utility from successive quantum hardware generations. Here we focus on one such algorithm: the accelerated variational quantum eigensolver (AVQE) \cite{Daochen}. Through the introduction of a parameter $\alpha\in[0,1]$ it is possible to interpolate between scaling regimes: decreasing the number of measurements in exchange for an increased circuit depth. The required number of measurements in AVQE can be given by:
\begin{equation}\label{eq:measurements}
N(p, \alpha) = \begin{cases}
\frac{2}{1-\alpha}\left(p^{-2(1-\alpha)}-1\right) & \mathrm{if}~\alpha\in[0,1)\\
4\log\left(p^{-1}\right)&\mathrm{if}~\alpha =1
\end{cases}.
\end{equation} Note that $N(p,0) =O(p^{-2})$ is the required number of measurements in VQE; $N(p,1) = O[\log(p^{-1})]$ is the required number of measurements in quantum phase estimation (QPE) up to further log factors. This is an up-to-exponential improvement over VQE in terms of measurements required for all $\alpha > 0$. As stated above this decrease in measurements is paired with a commensurate increase in circuit depth of $O(p^{-\alpha})$, offering an improvement over QPE for all $\alpha<1$. Equation~(\ref{eq:measurements}) is a decreasing function of $\alpha$; its behaviour is shown for various values of $p$ in Fig.~\ref{fig:measurements}.

\begin{figure}[t]
\centering
{
\setlength{\fboxsep}{5pt}
\setlength{\fboxrule}{1pt}
\fbox{\includegraphics[width=.95\columnwidth]{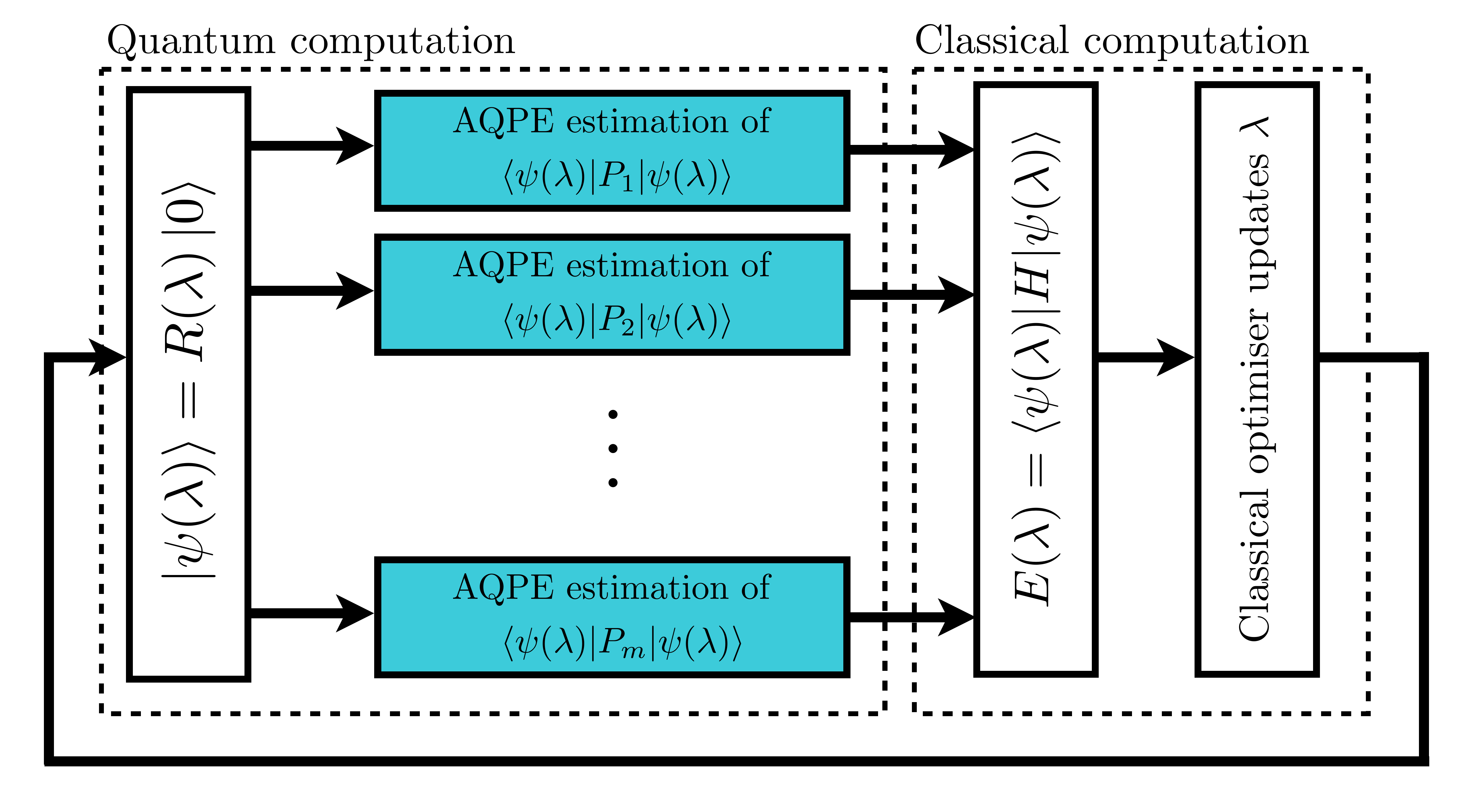}}
}
\caption{\label{fig:AVQE}Flow chart of the AVQE algorithm. }
\end{figure}

The task in question is minimisation of the energy of a given Hamiltonian, $H=\sum_ia_iP_i$ where $a_i$ are (known) complex coefficients and $P_i$ are Pauli matrices across subsystems of the Hamiltonian. Given an ansatz wavefunction $\ket{\psi(\lambda)}:=R(\lambda)\ket{0}$ parameterised by a real valued variable $\lambda$ and generated by a preparation circuit $R(\lambda)$, the energy of this Hamiltonian can be written:
\begin{equation}\label{eq:energy}
E(\lambda) = \sum_ia_i \Braket{\psi(\lambda)|P_i|\psi(\lambda)}. 
\end{equation}
 The process is to compile $R(\lambda)$ and pass $\Big\{\ket{\psi(\lambda)},\{P_i\}\Big\}$ to the QPU for each expectation value in Eq.~(\ref{eq:energy}) to be estimated. These are then collated and, using a CPU, the energy $E(\lambda)$ is calculated. This information is passed to a classical optimiser; the value of $\lambda$ is updated and the process begins again. Whereas VQE operates an algorithm known as quantum expectation estimation (QEE), AVQE replaces this with a modified Bayesian QPE algorithm known as AQPE. The generalisation from VQE to AVQE is then simply replacing this quantum subroutine. A flow-chart diagram of the algorithm is shown in Fig.~\ref{fig:AVQE}.

This modification has a consequence in the computational resources required. In contrast to VQE, the quantum subroutine in AVQE is adaptive, such that each circuit depends on the outcome of the previous. Consequently, a number of calculations are required between circuit evaluations. In the next section we will detail the action of this subroutine, before expanding on the precise steps needed to evaluate a single expectation estimation and exactly how this process could benefit from local computation.

\subsection{Quantum Subroutine}
The quantum subroutine in AVQE is the key driving factor around the advantages the algorithm exhibits over its competitors. It is a minor modification of an algorithm known as Rejection Filtering Phase Estimation \cite{bqpe}.
 Note that where VQE uses an algorithm to estimate expectations---which are the quantities needed to calculate the energy---we are instead implementing a phase estimation algorithm to indirectly achieve the value of the expectation. The justification for this is that through a careful choice of unitary applied to the target qubit, the output phase is directly proportional to $\Braket{\psi|P|\psi}$.

\begin{figure}[t]
\includegraphics[width=.99\columnwidth]{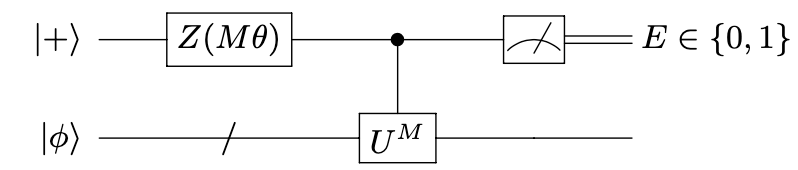}
\caption{\label{fig:aqpe}Phase estimation circuit. $M$, $\theta$ are free parameters.}
\end{figure}

 The circuit we need to implement is shown in Fig.~\ref{fig:aqpe}, with the goal to estimate the value of $\phi$ where $U\ket{\phi} = e^{i\phi}\ket{\phi}$, $\phi\in[-\pi,\pi]$. 
 Note this circuit has depth $O(M)$ and measurement outcomes are only recorded from the ancilla (control) qubit. The outcome probability $P\!\left(E\middle|\phi\right)$ of $E\in\{0,1\}$ is what gives us information on the eigenphase $\phi$. The rotation gate on the control qubit is $Z(M,\theta) = \mathrm{diag}(1, e^{-i M\theta})$ where parameters $M$, $\theta$ are free to be chosen.
 
 With the addition of a reflection operation $\Pi = \mathds{1} - 2\ket{0}\bra{0}$ we can specify the unitary $U = R\Pi R^\dag P R \Pi R^\dag P^\dag$ applied to input state $\ket{\phi}$. This operator defines a rotation by an angle $\phi$ in the plane spanned by $\{\ket{\psi},P\ket{\psi}\}$. It can be shown that $\phi = 2\arccos\big(\left| \Braket{\psi|P|\psi}\right| \big)$ which allows us to approximate   \[
 \left| \Braket{\psi|P|\psi}\right| = \cos\left(\phi/2\right).
 \]
While this process does not offer the sign of the expectation, this can be readily found through other methods.

This estimation of phase is statistical in nature, in that we assume the phase value $\phi$ follows a certain prior distribution with mean $\mu$ and standard deviation $\sigma$. Our goal, through repeated implementations of this circuit, is to update the distribution and reduce $\sigma$ below a certain threshold $p$, at which point we can take $\phi\approx\mu$ as our result.
Assume initially that the phase follows a normal distribution $f\sim\mathcal{N}(\mu, \sigma)$. Given our knowledge of the outcome probability $P(E|\phi)$ we can produce the posterior distribution $f(\phi|E)$ using Bayes' rule:
\[
f(\phi|E) = \frac{f(\phi) P(E|\phi)}{\int d\varphi f(\varphi) P(E|\varphi)}.
\]

To avoid directly calculating the posterior distribution, which is computationally hard, we instead  use a method known as rejection sampling \cite{wiebe2015bayesian}. We begin by collecting a measurement outcome $E\in\{0,1\}$ and sampled data $\Phi=\{\phi_j\}_{j=1}^{k}$ from the prior. By performing an acceptance test on each $\phi_j\in\Phi$ according to $P(E|\phi_j)$ we are able to `reject' certain values and the remaining (accepted) values are guaranteed to have been sampled from the true posterior. These accepted values then offer a new mean $\mu$ and standard deviation $\sigma$ for a new prior distribution; the update steps then continue until the requirement on $\sigma$ has been met. The cardinality $k$ of $\Phi$ is directly related to the reduction of $\sigma$ at each step; naturally for larger $k$ the approximation to the posterior mean and standard deviation is improved, but at the cost of performing more tests. 

\vspace{12pt}
 The next point to address is the specific form of outcome probability $P(E|\phi)$, which is related to the input state $\ket{\phi}$. We noted before that the unitary $U$ gives a direct relationship between the eigenphase and the desired expectation value. It can also be shown that $\ket{\psi}$, our prepared ansatz state, is in an equal superposition of the eigenstates of $U$:
\[\ket{\psi}=\frac{1}{\sqrt{2}}\left(\ket{\phi}+\ket{-\phi}\right),\]meaning that $\ket{\psi}$ must be collapsed into one of $\ket{\pm\phi}$ prior to running the circuit. Collapsing into one of $\ket{\pm\phi}$ gives an outcome probability:
\begin{equation}\label{eq:collapsed}
P\big(E|\pm\phi\big)=\frac{1}{2}\Bigg(1+(1-2E)\cos\left[M\left(\theta\mp\phi\right)\right]\Bigg).
\end{equation} For details on the collapse process please see Ref.~\cite{Daochen}, Appendix C.

The parameters $M$, $\theta$ each play an important role in the algorithm. The value of $M$ sets the overall depth of the circuit but also enters as a multiplicative factor in Eq.~(\ref{eq:collapsed});  the parameter $\theta$ acts as a rotation that, with the correct setting, can offset the rapid oscillations induced by $M$. The authors of Ref.~\cite{bqpe} chose $M=\lceil1.25/\sigma\rceil$ meaning that upon each update on your posterior distribution (i.e. defining a new $\mu$, $\sigma$) the circuit depth changes. For a precision $p$, the final circuit depth would be $O(p^{-1})$ --- the circuit depth regime of QPE. We maintain this dynamic behaviour of circuit depth but instead set \[M=\mathrm{max}\left(1,~ \left\lfloor \frac{1}{\sigma^\alpha}+\frac 1 2\right\rfloor \ \right)\]for parameter $\alpha\in[0,1]$. 
By pre-setting the value of $\alpha$ we are able to control how quickly the circuit depth grows and to what maximum it reaches, $O(p^{-\alpha})$. Finally, we set the parameter $\theta = \mu - \sigma$ as in Ref.~\cite{Daochen}.

\begin{figure}[t]
\hspace{-20pt}\includegraphics[width=.95\columnwidth]{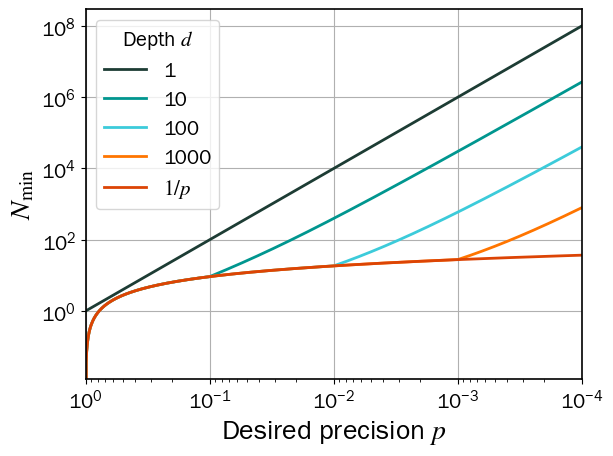}
\caption{
\label{fig:nmin}
Minimum number of measurements required $N_\mathrm{min}$, Eq.~(\ref{eq:minmeas}), against the target precision $p$ plotted here for various coherent circuit depths $d$. Setting $d=1$ recovers the scaling of VQE, and $d = 1/p$ is the scaling regime of QPE.
}
\end{figure}
\subsection{Performance \& Applications for Local Computation}
As the behaviour of the classical outer-loop in AVQE is problem-dependent and benefits from various optimisation protocols, the performance of AVQE is largely centered around the action of the quantum subroutine. Put another way, the accuracy in estimating the ground state energy of a Hamiltonian by AVQE is dependent on the accuracy of AQPE in measuring the eigenphase $\phi$.

The adaptive nature of the subroutine means it requires a large number of classical calculations to be completed before a single eigenphase can be estimated. This is in contrast to VQE which benefits from batching, i.e. pre-computing a series of circuits and passing these to the QPU to be executed in order. The individual steps for an eigenphase calculation in AVQE are: sampling $k$ values of $\phi$ from a normal distribution with mean $\mu$ and standard deviation $\sigma$, testing each sampled $\phi$ against Eq.~(\ref{eq:collapsed}) and storing accepted values, and finally calculating a new $\mu$ and $\sigma$ to inform a new prior. 
These steps need to be repeated multiple times to achieve a value of $\phi$. The exact number of iterations (equivalently, measurements) is primarily related to the desired precision $p$ and acceleration parameter $\alpha$, but we are further restricted through hardware specifications. For a given coherent circuit depth $d$ the optimal value of $\alpha$ is found by minimising the number of measurements Eq.~(\ref{eq:measurements}). For simplicity let us set $d = p^{-\alpha}$, $p\ll1$, such that the maximum value of $\alpha$ is given by \[\alpha_{\mathrm{max}}=\mathrm{min}\left\{-\frac{\log(d)}{\log(p)},~1\right\}.\]
\begin{figure}[t]
\hspace{-20pt}\includegraphics[width=.95\columnwidth]{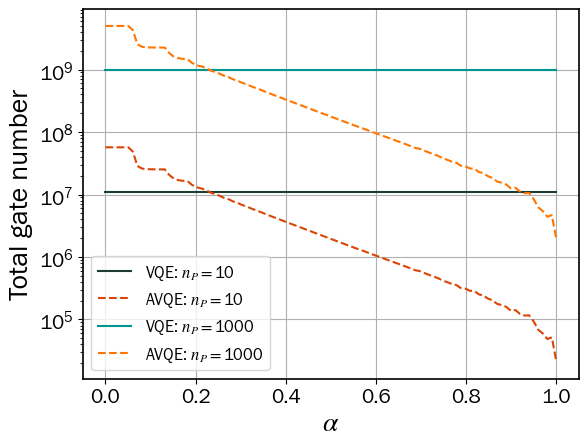}
\caption{
\label{fig:AVQE_v_vqe}
Total number of gates required to reach precision $p=10^{-3}$ shown for VQE (solid) and AVQE (dashed). Two preparation circuit lengths are reported, $n_P=10$ (blue, orange) and $n_P = 10^3$ (green, red).
}
\end{figure}
The minimum number of measurements required to achieve a precision $p$ is therefore
\begin{equation}\label{eq:minmeas}
N_{\mathrm{min}}(p, d) = 
\begin{cases}
\frac{2 \log(p)}{\log(pd)}\left(\frac{1}{p^2d^2}-1\right) & pd<1\\
4\log(p^{-1}) &pd > 1
\end{cases}.
\end{equation}
Figure~\ref{fig:nmin} shows the behaviour of Eq.~(\ref{eq:minmeas}) for decreasing value of precision, and various maximum circuit depths $d$. For comparison purposes the circuit depth of VQE (blue) and QPE (purple) are shown. The strength of AVQE lies in this range with the `acceleration' epitomised by the up-to-exponential decrease in required measurements. We propose that AVQE is a candidate for utilising the full potential of near-term NISQ devices, and, in particular, that its quantum phase estimation subroutine can make effective use of  local computation. 

While overall AVQE requires more computational effort, a further advantage over traditional VQE lies in reducing the required quantum resources, specifically in the number of gates that need to be implemented. VQE operates with a fixed circuit depth, that of $n_P + 1$ where $n_P$ is the gate length of the preparation circuit. Comparatively, AVQE requires much deeper circuits and these continue to increase as the algorithm runs. With this in mind, for $\alpha = 0$ ($M\equiv1$) there are limited benefits to running AVQE in comparison to VQE. For $\alpha > 0$ however, the decrease in number of measurements Eq.~(\ref{eq:measurements}) begins to outweigh this increased computational cost. To illustrate this trade-off we can consider the number of gates in the preparation circuit, the common factor between both algorithms.

To achieve a precision $p$ in VQE, the total gates required is (approximately) $(n_P + 1)p^{-2}$.
For AVQE a single execution of the circuit in Fig.~\ref{fig:aqpe}, ignoring possible parallelisation, requires $4M(n_P+1)+n_P+3$ gates. The dynamic nature of $M$ makes the calculation of total gates required to achieve a precision $p$ difficult, however we can numerically approximate its schedule. The comparison between these two is shown in Fig.~\ref{fig:AVQE_v_vqe} for a precision $p = 10^{-3}$, equivalently a fidelity of $99.9\%$. We report two extreme values of $n_P$: 10 and $10^3$ with VQE shown in solid lines (blue, green for $n_P=10$, $10^3$ respectively) and AVQE shown in dashed (orange, red). For both values of $n_P$ the intersection happens at $\alpha \approx 0.23$. That is to say, beyond this value AVQE requires fewer quantum computational resources than VQE.
The plateau at the beginning of the AVQE lines are a result of the low values of $\alpha$--- the value of $M$ never increases beyond 1. The second plateau is the case in which $M$ never increases beyond 2; as $\alpha$ increases the acceleration of $M$ causes these plateaus to disappear.


AVQE exhibits clear advantages over VQE, however, the intermediate calculations can still incur a runtime penalty. In an effort to reduce the overall wall-clock time as many as possible of these calculations can be done locally. By relegating the CPU so that it only operates on the classical outer-loop of the algorithm we can utilise the abilities of the lower levels in the computational stack for subroutine calculations. If we were to continually communicate back to the CPU after each recorded measurement outcome, the latency between the QPU and CPU would create untenable runtime. In VQE (and others) the relevant sequence of circuits to be run can be pre-computed. However, it is not possible in this case, as circuit parameters $M$, $\theta$ depend on the outcome of the previous circuit.


\begin{figure}[t]
\hspace{-20pt}\includegraphics[width=.95\columnwidth]{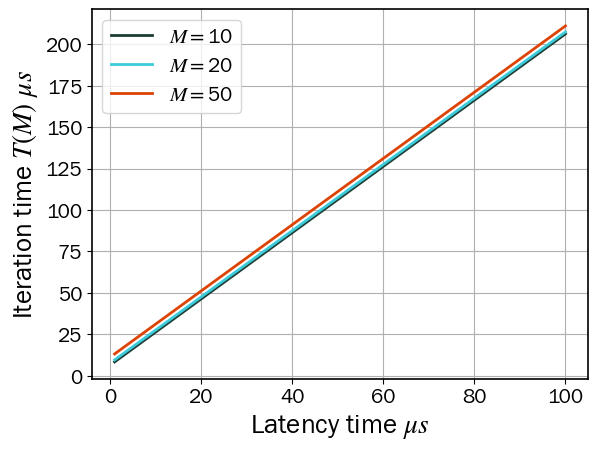}
\caption{
\label{fig:scq}
Algorithm iteration of AVQE for a generic $M$-depth circuit consisting of 2-qubit gates only using approximate hardware times for superconducting qubits implementations.
}
\end{figure}

\subsection{Costing the subroutine}\label{sec::cost_subroutnine}
To emphasise the benefits of implementing this algorithm in a low-latency environment we here provide a comprehensive view of the computational cost of AVQE. For ease of discussion we introduce the following parameters: $t_{lat}$ denotes the latency between QPU and the classical computation block (either CPU or alternative); $t_c(M)$ is the time taken to implement the circuit in Fig.~\ref{fig:aqpe} for a certain depth $d\approx M$, including the circuit to prepare the input state $\ket{\phi}$.
The qubit reset and measurement times are denoted $t_r$, $t_{meas}$ respectively; as these can be implemented simultaneously we need only consider $\mathrm{max}(t_r,~ t_{meas})$. Finally, the time taken to perform the Bayesian update on the probability distribution is then $t_B$. Note that the latency time $t_{lat}$ needs to be included twice: once for communicating the circuit instructions to the QPU and again for the measurement outcome being sent back. One iteration of AQPE has the following cost then:
\begin{equation}\label{eq:T}
T(M) = 2t_{lat}+t_c(M)+ \mathrm{max}(t_r,t_{meas})+t_B,
\end{equation}
and the cost of estimating the eigenphase: 
\[
\tau = \sum_M a_M(p, \alpha)T(M)    .
\]The real-valued coefficients $a_M(p, \alpha)$, which determine how many circuit evaluations take place for a value $M$, represent the only unknowns in this costing. Its dependencies are due to the fact $\alpha$ and $p$ both play a key role in the schedule of $M$; the process is probabilistic due to the nature of the Bayesian update but will manifest as a step function over $\mathds{Z}^+$. We will not address methods for investigating the behaviour of $a_M$ here as we believe it is outside the scope of this document.

We will evaluate the time taken for a single iteration Eq.~(\ref{eq:T}) in both a superconducting qubit and trapped ion implementation. As the circuit being implemented is dominated by $M$, for simplicity we will consider circuits of depth $M$ consisting only of 2-qubit gates for each hardware specification. The Bayesian update time $t_B$ is performed outside of the quantum hardware; an optimised and hardware-conscious rejection sampling calculation can be performed in approximately 5$\upmu$s.

\begin{figure}[t]
\hspace{-20pt}\includegraphics[width=.95\columnwidth]{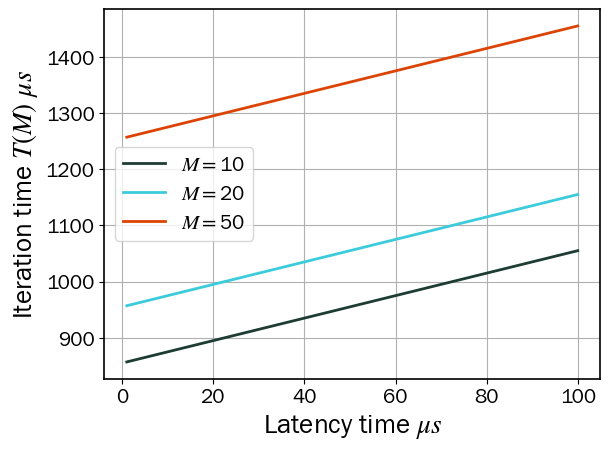}
\caption{
\label{fig:ti}
Algorithm iteration of AVQE for a generic $M$-depth circuit consisting of 2-qubit gates only using approximate hardware times for trapped ion implementations.
}
\end{figure}

For superconducting qubit hardware the readout and reset of qubits can be done in 120ns, with a similar 2-qubit gate time. For trapped ions the times are much slower, with a 2-qubit gate time of 10$\upmu$s and readout and reset time of approximately 750$\upmu$s. We can see the result of the superconducting simulation in Fig.~\ref{fig:scq} with a theoretical latency range of $[1, 100]\upmu$s. Due to the very low gate times of this technology the benefit of a lower latency in the calculation is stark; up to a 20x improvement on runtime can be achieved. Comparatively the improvement for trapped ion technology, Fig.~\ref{fig:ti}, is much more slight due to the long gate times. However we argue that an improvement in runtime, however small, is advantageous in NISQ hardware.

\subsection{Closing remarks}
The accelerated variational quantum eigensolver algorithm is able to outperform traditional VQE in terms of measurements, utilisation of quantum coherence in hardware and the overall required quantum resource budget. These advantages are coupled with the requirement for regular, generally non-trivial, classical calculations and low batch size. With this in mind, operating AVQE in a low latency environment can accentuate its performance and allow for more efficient implementations.

The future of NISQ hardware and showcasing of quantum advantage on these devices is, we believe, rooted in these kinds of adaptive algorithms. While AVQE is not unique in incorporating coherent circuit depth into algorithm functionality (references \emph{passim}), it is a prime example of the types of intermediary calculations adaptive algorithms will require. Efficient implementation of these algorithms can be achieved through the use of a degree of local computation. 


\section{Quantum Error Correction}\label{sect:qec}
\subsection{Introduction to QEC}
An area where local computation is likely to prove advantageous is error correction. In the fault tolerant regime, error correcting protocols are going to be necessary to achieve the required coherence times. However, even in the NISQ era, 
some low level error correction is likely to be useful \cite{holmes}. 

Error correction on a quantum computer is performed by a repetitive sequence of events. First, a bit string called a \emph{syndrome} is created. This
is a record of the error that has occurred on the data, and on a quantum computer the syndrome is created by entangling ancilla qubits to the data qubits and measuring certain Pauli operators corresponding to the code being used. The syndrome then needs to be decoded by a classical decoder, which outputs the best guess of the error that has occurred. 
This information then needs to be relayed back to the QPU in order for the
error to be taken into account, either via a correction or an update to the set of gates that are subsequently going to be applied.

To prevent the build up of errors this error correction cycle needs to be performed at regular intervals. If the decoder is situated on the CPU, latency will become an issue given the frequency of this task. 
Moreover, given the number of operations required to perform an error correction cycle,
and the frequency with which they need to be transmitted, bandwidth will also be an issue.
However, we show in this section that if the decoder can be implemented using local computation then the latency and bandwidth bottlenecks can be overcome.

\subsection{The Surface Code}
\begin{figure}[ht]
\includegraphics[scale=0.16]{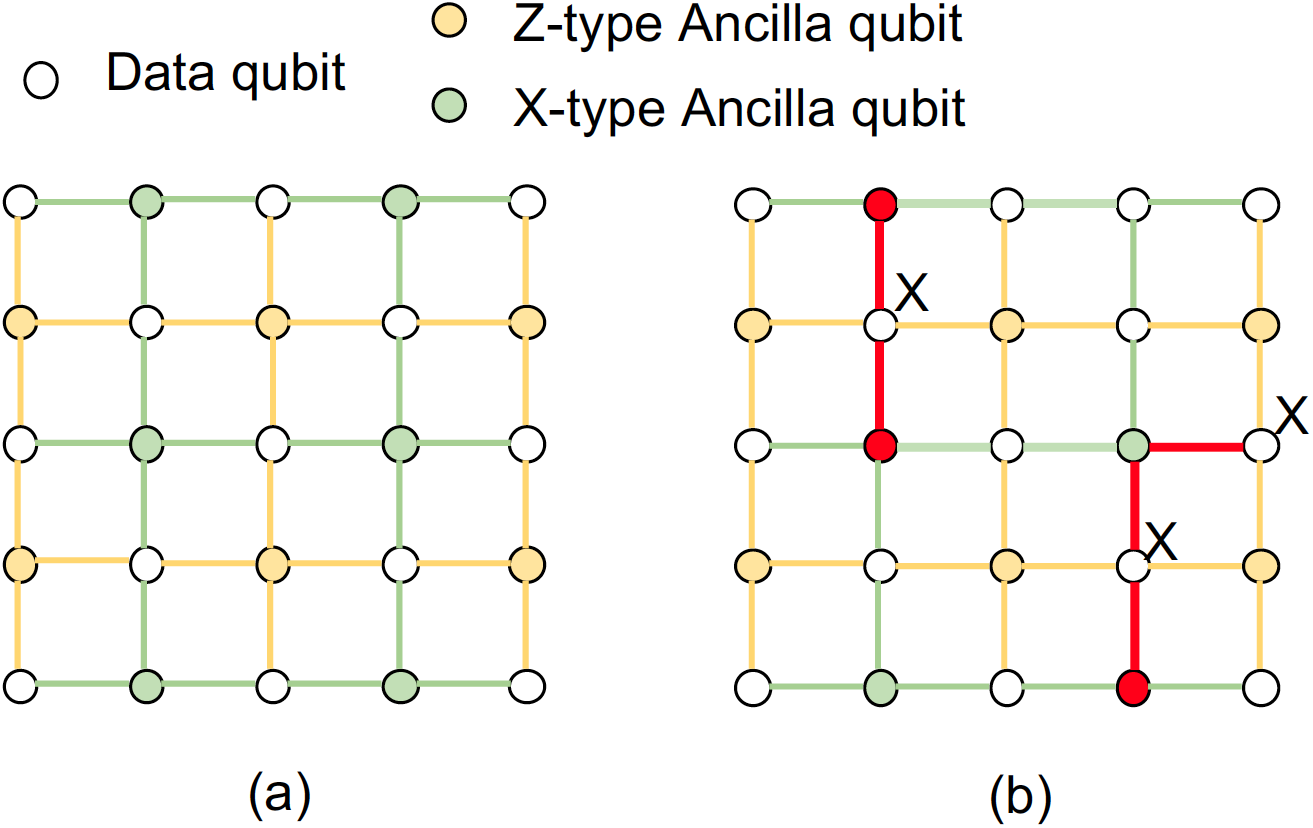}
\caption{The grid of qubits for the distance $3$ surface code, taken from \cite{delfosse}. The $X$ ($Z$) ancilla qubits are used to measure $X$-errors ($Z$-errors). The red edges represent $X$ errors on the corresponding data qubits.}\label{fig:surface}
\end{figure}

The surface code is the most promising family of quantum error correcting codes due to its high noise threshold, which means that they can handle a very noisy error rate.
The surface code considered in \cite{delfosse} is laid out on a $(2d-1)\times (2d-1)$ square grid of qubits (see Figure \ref{fig:surface}). The qubits consist of data qubits, which store the logical information, and ancilla qubits,
which are used to measure the syndrome. Such a code has distance $d$, which is a measure of its error tolerance.

The fact that the surface code can be laid out on a two-dimensional grid means that the ancilla qubits only need to interact with their nearest neighbour data qubits. After a syndrome measurement, an ancilla qubit ``lights up" if the parity of the measurements with its neighbour data qubits is odd. We shall call such a qubit a \emph{hot syndrome qubit}, or just a \emph{hot qubit}. The syndrome
produced by a syndrome measurement is a bit string that records the set of hot syndrome qubits.

The circuit used to measure the syndrome is susceptible to the same noise that affects the data qubits. A consequence of this is that measurement errors can cause an incorrect syndrome to be produced, as well as introduce errors
to the data qubits. In order to overcome this problem, several rounds of syndrome extraction are performed,  in particular, $d$ rounds are performed in 
the case of a distance $d$ surface code. From the accumulated information contained in these multiple syndromes, a good decoder will output an error which best explains them.

\subsubsection{The Union-Find decoder}
\begin{figure}[ht]
\includegraphics[scale=0.16]{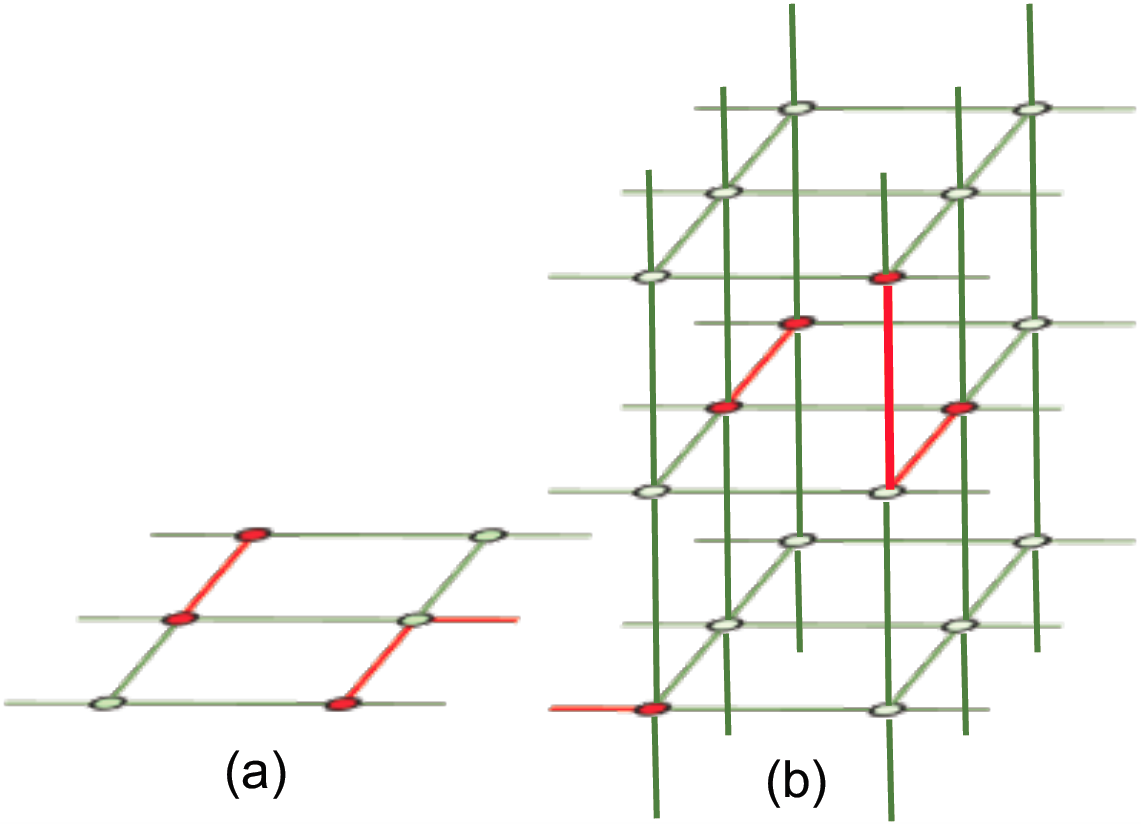}
\caption{The graph in (a) is a decoder graph for a single round of syndrome measurements. The graph (b) is is a three-dimensional decoder graph for three rounds of syndrome measurement. The horizontal red edges represent $X$ errors on qubits; the vertical red edges represent syndrome bit errors. Figure taken from \cite{delfosse}.}\label{fig:decodergraph}
\end{figure}

Many decoding algorithms for the surface code use the \emph{decoding graph} as their input. The decoding graph captures the results from the $d$ rounds of syndrome measurement 
and it is best represented as a three-dimensional graph on a lattice of vertices consisting of $d$ layers with edges connecting nearest neighbours. 
Each layer of the lattice records the results of a single round of syndrome extraction; the vertices represent the ancilla qubits and each horizontal edge in a single layer represents a possible error that can occur on the corresponding qubit. 
Each vertical edge between layers represents a possible syndrome bit flip error (see Figure \ref{fig:decodergraph}). The vertices corresponding to the hot syndrome qubits after the $d$-rounds of measurement
are highlighted in the graph (coloured red). The goal of any decoding algorithm is to come up with an error pattern consisting of both data qubit errors and syndrome bit errors 
that best describes the set of hot syndrome qubits in the decoding graph.

\begin{figure}[ht]
\includegraphics[scale=0.30]{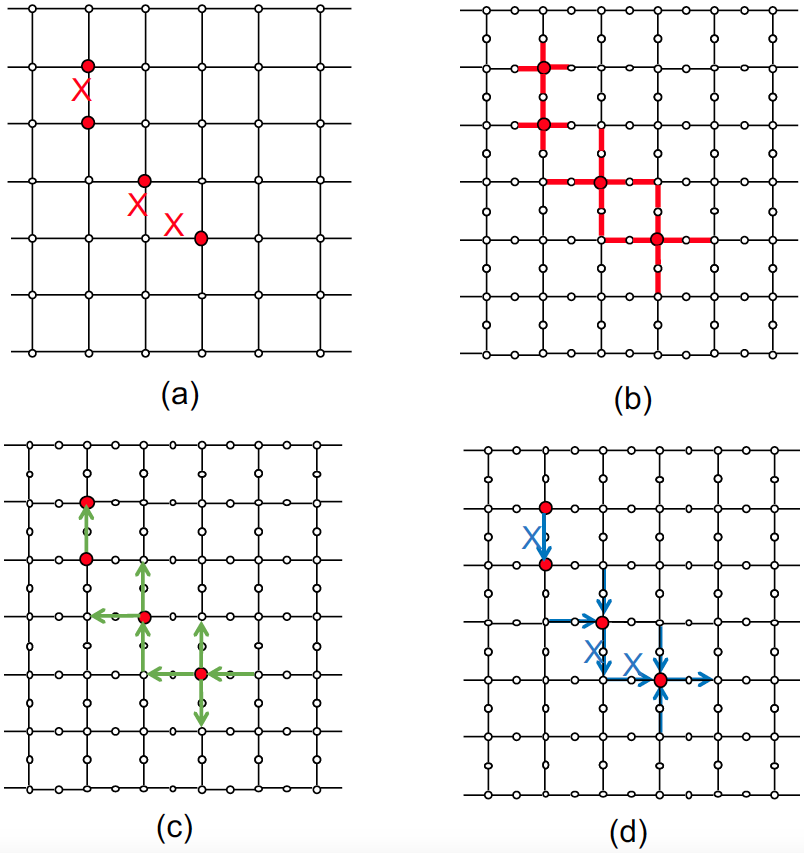}
\caption{The four stages of the Union-Find decoder, taken from \cite{delfosse}.}\label{fig:peeldec}
\end{figure}

The Union-Find algorithm \cite{unionfind, peeldec} is an algorithm that performs this task in almost linear time.
We illustrate how the algorithm works in Figure \ref{fig:peeldec}, for only one level of the decoding graph, that is, assuming there are no measurement errors.
First in (a), the syndrome is measured giving a set of hot syndrome qubits (these are the red qubits). 
Once the syndrome has been measured, clusters of edges are grown around the hot syndrome qubits by expanding along half edges of the lattice in all directions, which is shown in (b).
During this process, clusters that overlap are combined and if at any point a cluster contains an even number of hot syndrome qubits, it stops growing. 
The cluster growth process continues until all clusters contain an even number of hot syndrome qubits.
Once the cluster growth process ceases, a spanning tree for each cluster is created, shown in (c). Finally in (d), given the spanning trees for the clusters, the Peeling decoder is applied to each spanning tree to decode the error.


\subsection{Decoding must be fast}

Any quantum circuit that cannot be simulated classically must contain non-Clifford gates \cite{gott}, and in order to apply a non-Clifford gate it is necessary to know 
the current state of the errors. This is because, assuming stochastic Pauli noise, unlike for Clifford gates, a general error cannot be handled by commuting it through the non-Clifford gates
where it can be dealt with at a later time. Therefore a decoder must keep up with the rate at which syndromes are being generated, otherwise a data backlog will be created.

To clarify this, consider the ratio $f$ of the rate of syndrome generation $(r_{gen})$ versus the rate of decoder processing $(r_{proc})$.
In \cite{holmes} it is shown that the data backlog caused by a slow decoder will lead to a latency overhead that grows exponentially in $f$, namely if a circuit contains $k$ non-Clifford gates
then the latency scales as $f^k$. So any protocol with $f>1$ is going to be problematic. 

To see the impact of this, consider the circuit given in \cite{holmes2} designed to perform a multiply-controlled $CNOT$ gate on $100$ logical qubits. It consists of $\sim 2356$ gates, of which $686$ are $T$-gates (which are non-Clifford). With the caveats that the syndrome generation cycle time is approximately $400$ ns \cite{gosh}, and the decoder requires $800$ ns to execute \cite{chamberland}, 
the ratio $f=2$ leads to an execution time of approximately $10^{196}$ seconds (orders of magnitude longer than the age of the Universe). This motivates us to consider fast decoding protocols.

\subsection{A micro-architecture for the surface code} 

As commented on above, any classical decoder must finish before the following rounds of syndrome measurement have completed. Until recently it was not known whether such a decoder for the surface code existed \cite{fowler1}. However Das et al.~\cite{delfosse} have proposed a micro-architecture to implement the Union-Find decoder for the surface code that overcomes this hurdle. This micro-architecture deals with the three stages of the
Union-Find decoder that can be implemented classically. These are the growing of the clusters, the creation of the spanning trees, and the application of the Peeling decoder. 

\subsubsection{Growing the clusters}

In the micro-architecture the Graph-Genenerator (Gr-Gen) engine 
creates the clusters around the syndrome vertices. 
It consists of a Spanning Tree Memory (STM), a Zero Data Register (ZDR), a Fusion Edge Stack (FES), a Root Table, a Size Table, a
Parity Register and a Tree Traversal Register.

The STM keeps track of the clusters - it stores one bit for each vertex and two for each edge. 
The ZDR is used to quickly look up the STM; each entry of this register corresponds to a row of the STM and if a row in the STM contains a non-zero bit, then the corresponding register bit is $1$, otherwise it is equal to $0$.
Newly grown edges are stored in the FES, from where they can be added to the existing clusters in the STM. The root and size tables respectively keep track of the roots and sizes of each cluster.
The parity register keeps track of the parity of each cluster; it stores a $1$ if the corresponding cluster is odd, that is, the cluster contains an odd number of hot syndrome qubits.

The Gr-Gen grows clusters by first reading the parity register to identify odd clusters. Using the information stored in the root and size tables, as well as the FES, 
it then grows the odd clusters by reading and writing to the STM. Newly added edges are checked to see if they connect two clusters. 
This is done by checking the root of each vertex incident with an edge (note, this process is aided by the Tree Traversal Register). If
the two roots are different, the corresponding clusters are subsequently merged. This involves updating the root and size tables. 

\subsubsection{Creating the Spanning Trees}

The spanning trees, which are used as the input to the peeling decoder, are created by the Depth First Search (DFS) Engine. In the DFS engine a depth first search algorithm is applied to the clusters stored in the STM,
and it is implemented using a finite state machine and two edge stacks. The reason edge stacks are used is that the implementation of the peeling decoder requires the edges in a spanning tree to be
traversed in reverse order. Moreover, two stacks are used to enhance performance; while a spanning tree is created in one edge stack by the DFS engine, the peeling decoder can be applied to 
a spanning tree for a cluster stored in the other edge stack.

\subsubsection{Implementing the Peeling Decoder}

The Correction engine (Corr engine) implements the peeling decoder described in \cite{peeldec}. It needs access to the spanning trees held on the edge stacks in the DFS engine as well as the bits corresponding to the hot syndrome qubits in the STM. To reduce latency, these syndrome bits are saved to the edge stack when the spanning trees are created so that the Corr engine only needs to read the edge stack.
As part of the peeling decoder, the syndrome is dynamically changed. A temporary syndrome register keeps track of this in the Corr engine. Finally, the result of the peeling decoder is recorded by updating the Error Log. 
If the error to be recorded cancels out an error held in the Error Log from a previous correction cycle, the Error Log is updated to reflect this.

\subsubsection{Performance}

By running simulations, it is observed in \cite{delfosse} that the Gr-Gen engine takes roughly twice as long as both the DFS engine and the Corr engine. If one decoding block consists of
one of each type of engine, then the DFS engine and Corr engine will spend a lot of time waiting for the Gr-Gen engine to finish. This suggests that a more efficient configuration can be achieved. 
A decoding block consisting of 2 Gr-Gen engines, one DFS and one Corr engine is proposed in \cite{delfosse}, which processes 2 logical qubits. 
Such a decoding block is labelled a $(4,2,1,1)$-decoding block, where the $4$ corresponds to the number of logical $X$ and $Z$ operators that are being protected. Using these decoding blocks, an algorithm that requires $L$ logical qubits will require $L$ Gr-Gen engines, $L/2$ DFS engines and $L/2$ Corr engines.

\begin{figure}[ht]
\includegraphics[scale=0.16]{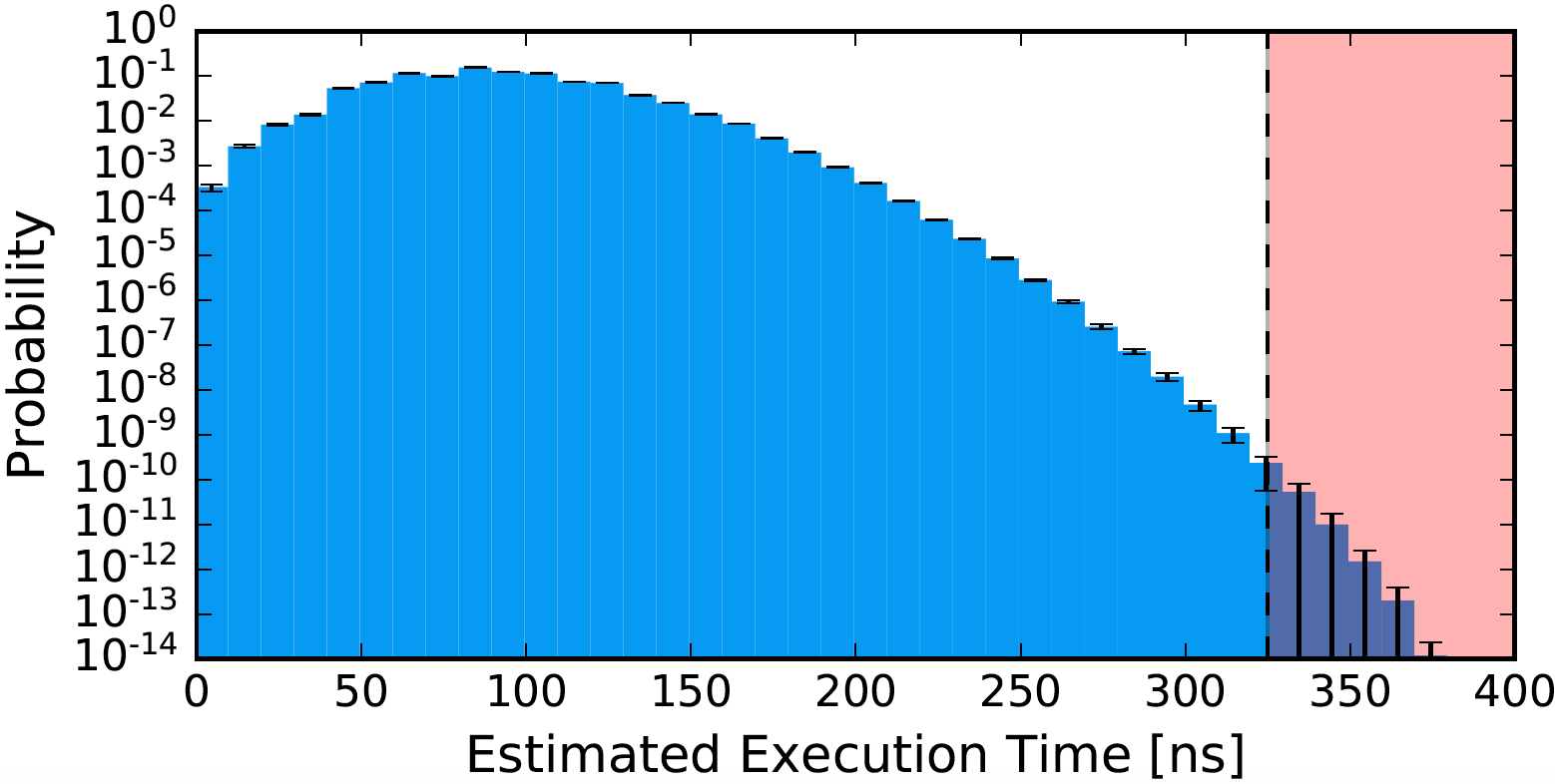}
\caption{A Monte-Carlo simulation of the micro-architecture implementing the $(4,2,1,1)$-decoding block, taken from \cite{delfosse}.}\label{fig:peelperform}
\end{figure}

Now recall that when using a distance $d$ surface code, $d$ rounds of syndrome extraction are performed in one correction cycle. We call this a \emph{complete measurement cycle}. In order for
the decoder to keep up with error correction, it must finish its decoding before the subsequent complete measurement cycle is finished, because otherwise errors might accumulate and
spread in an uncontrollable manner. If the decoder does not finish before the subsequent complete measurement cycle, we call this a \emph{timeout failure}. 
One way that the impact of timeout failures can be reduced is to ensure that they are less likely than logical errors. That is, the probability of a timeout failure of a decoder block per logical qubit must be less than 
the probability of a logical failure. Using the $(4,2,1,1)$-decoder block, this implies that
$$p_{ToE}(d,p)/2\leq p_{Log}(d,p),$$
where $p_{ToE}(d,p)$ is the probability of timeout failure 
in a decoder block and $p_{Log}(d,p)$ is the logical failure rate.
Implementing the Union-Find decoder on the micro-architecture outlined above facilitates this requirement as it reduces latency. Specifically, assuming a surface code with distance $d=11$ is being used on a $(4,2,1,1)$-decoding block 
with a physical error rate of $10^{-3}$, then a block of $2$-logical qubits can be decoded in $325$ $ns$ (see Figure \ref{fig:peelperform}).
This is well below the time it takes to perform a complete measurement cycle for a surface code with these parameters, which is approximately $11$ $\mu s$ \cite{gidney}.

\subsection{NISQ+ Regime}

The micro-architecture described in the previous section is designed for 
fault tolerant machines. In contrast, we now show the benefits of local computation
on near term devices. 

In \cite{holmes} the authors propose an approximate decoding algorithm to overcome the latency overhead problem. The important point to make here is that they sacrifice
some accuracy in their decoder in order to be efficient. The classical decoder is implemented using single flux quantum (SFQ) logic; is a classical logic implemented in superconducting hardware. The decoder design is built out of a 2d array of modules implemented in SFQ logic circuits;
the 2d array represents the qubits of the 2d surface code. Specifically, each module represents either a data qubit or an ancilla qubit. 

The decoding algorithm proceeds by first identifying the two modules representing the two hot syndrome qubits that are closest together.
Next, a chain of modules representing a path connecting these two qubits is recorded. Finally, the two chosen modules
are reset and the algorithm iterates on to the next closest pair of hot qubits. The first part of the algorithm that identifies the two closest hot qubits is performed by implementing the cluster growth stage of the
Union-Find decoder. Moreover, the modules are hardwired so that only certain paths between qubits can be recorded, leading to
an approximation of an ideal decoder.

One obvious problem with the algorithm as outlined above occurs when there are multiple hot qubits that are equidistant. To overcome this, the authors introduce a request-grant policy that allows the hardware to select a subset of pairs of the equidistant hot qubits. Similarly, a hardware solution is provided for hot qubits that are close to the boundary of the surface code lattice.
    
\subsection{Performance of the SFQ logic decoder}
\begin{figure}[h]
\includegraphics[scale=0.19]{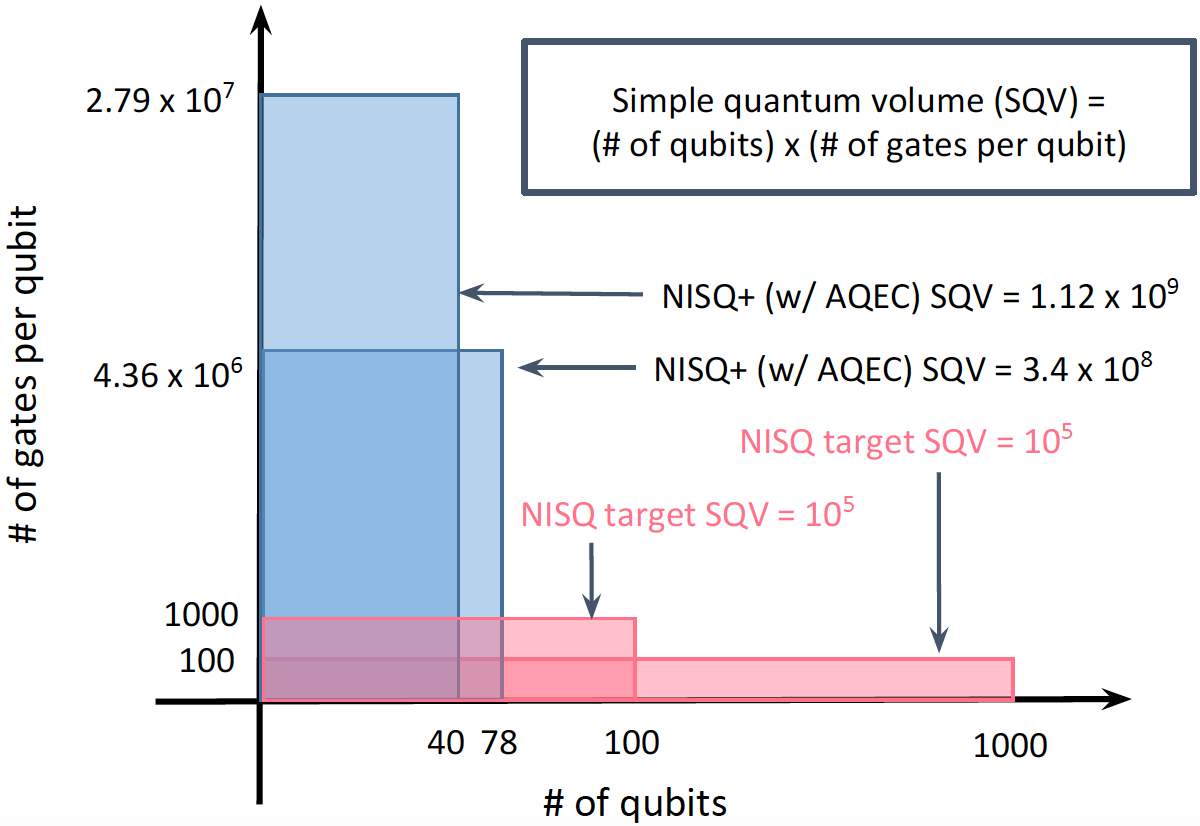}
\caption{Simple quantum volume of a near term device with 1000 physical qubits
and an error rate of $10^{-5}$, taken from \cite{holmes}.}\label{fig:nisq2}
\end{figure}

One of the metrics used in \cite{holmes} to evaluate their decoding protocol is simple quantum volume (SQV). This can be defined as the product of the number of computational qubits of a machine by the number of gates 
expected to be able to perform without error. It is shown that in near term machines (termed NISQ+ machines in \cite{holmes}), their proposed decoding protocol can increase the SQV.
In particular, the authors consider a device of 1000 physical qubits with an error rate of $10^{-5}$, an extension of a machine that is predicted to exist in the near term \cite{ibm}. Using a surface code with distance $3$ which encodes 78 logical qubits, 
the authors show that their protocol
can increase the SQV from $10^5$ to $3.4\times 10^8$. Similarly a code of distance $5$ that
encodes 40 logical qubits increases the SQV to $1.12\times 10^9$ (see Figure \ref{fig:nisq2}). 

\begin{figure}[h]
\includegraphics[scale=0.18]{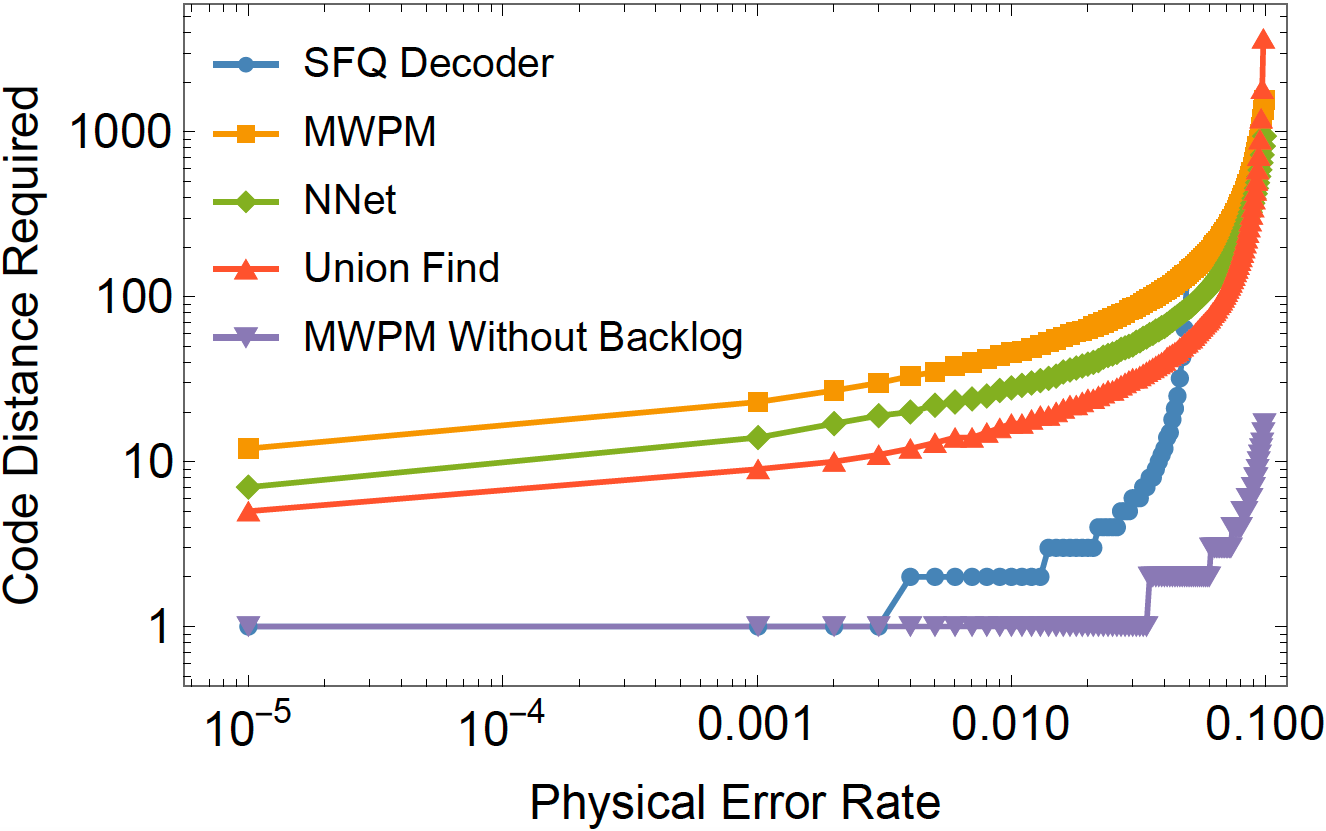}
\caption{Comparison of required code distances of different decoders to
execute an algorithm consisting of 100 T-gates, taken from \cite{holmes}. Compared are the SFQ
Decoder, minimum weight perfect matching decoder (MWPM) \cite{fowler2}, neural
network decoder \cite{bair}, union find decoder \cite{unionfind}, 
and a theoretical MWPM decoder with no backlog, across both code distances 
and physical error rates.}\label{fig:nisq1}
\end{figure}

Despite the approximate nature of the SFQ logic decoder, the
reduction in resources due to its speed
compares favourably to other decoding protocols, as can be seen in Figure \ref{fig:nisq1}.
This is in part because, if the ratio is $f>1$, the backlog in the bottleneck increases the effective logical error rate as many more syndrome cycles are needed to process one logical gate. Therefore it is clear that one needs to take into account both the speed of the decoder
and the latency overhead when assessing the efficiency of a decoding protocol. 

\subsection{The instruction bandwidth problem}
In \cite{tannu} a bottleneck caused by the instruction bandwidth for quantum error correction (QEC) is considered. Any fault tolerant device will require QEC, and it has been proposed that this should be managed at the software level. The reason for this is twofold. First, the optimum quantum error correcting protocol 
for a quantum device is dependent on the hardware qubit properties as well as the algorithm that is going to be implemented. Therefore it is suggested that a QEC library will be necessary in order to minimise the resources 
to run a particular algorithm \cite{svore}. Second, quantum error correction is an active research area, and so one would want to add to the QEC library as more efficient codes are discovered.

\begin{figure}[ht]
\includegraphics[scale=0.16]{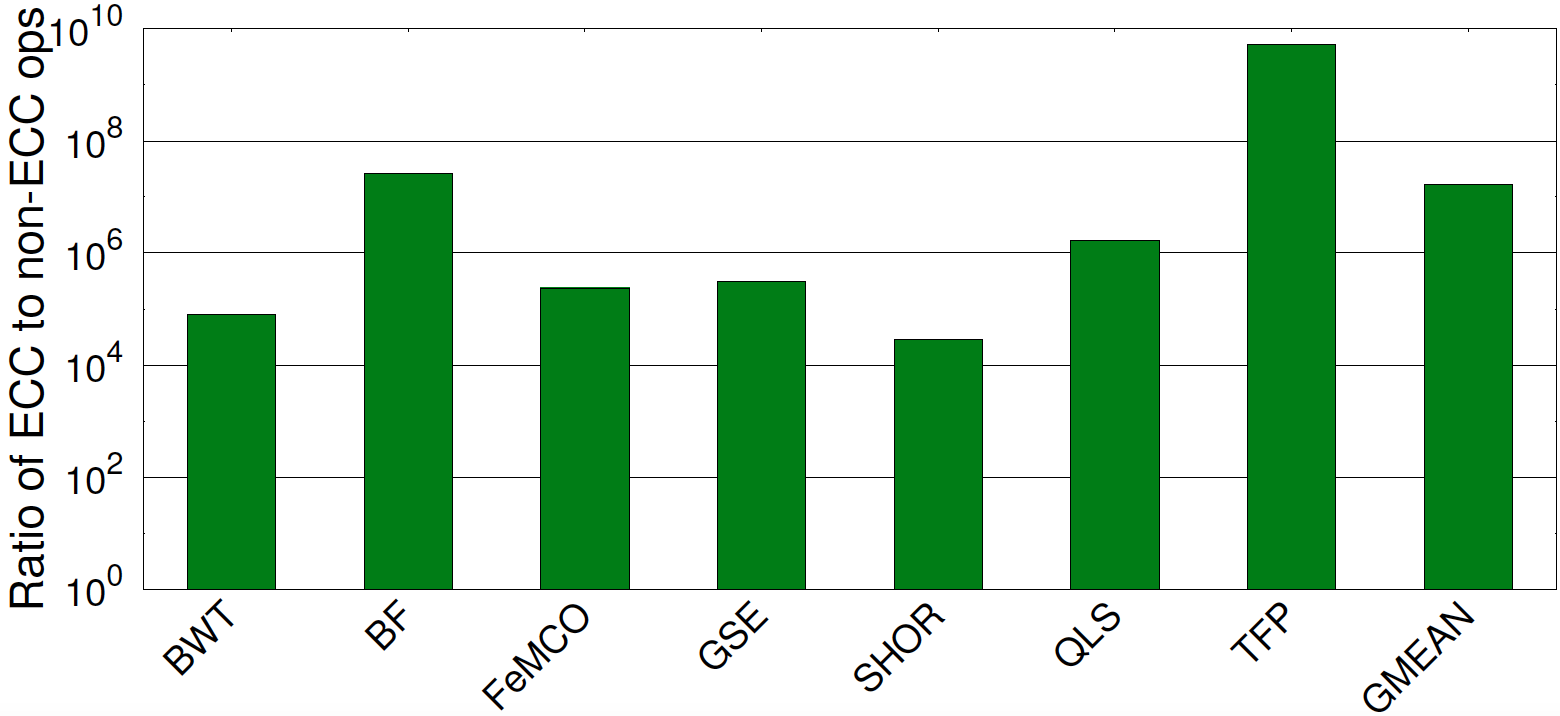}
\caption{Ratio of QEC operations to non-QEC operations for seven quantum algorithms, taken from \cite{tannu}.}\label{fig:qecband}
\end{figure}

However, controlling the QEC at the software level requires the QEC instructions to be transmitted on the same channel as the program instructions. This leads to an instruction bottleneck.
Specifically, in \cite{tannu}, the quantum resource estimator toolbox (QuRE) \cite{qure} is used to cost the instruction bandwidth for error correction for seven quantum algorithms. They find that at least $99.99\%$ of the total instruction bandwidth is taken up by the QEC instructions
when the error correction is managed at the software level (see Figure \ref{fig:qecband}). 
To demonstrate the problems that this can cause, consider a device comprised of superconducting qubits that operate at 100 MHz with byte sized quantum instructions. 
In order for a qubit to maintain its integrity it must receive QEC instructions at roughly its operating rate. 
Therefore each qubit requires 100 MB/s of QEC instruction. However, this implies that a quantum computer with 100,000 qubits would require 10TB/s of QEC instruction bandwidth.

Large instruction bandwidth can be handled in traditional computing systems by caching instructions.
However, this can lead to small instruction delays which are not acceptable for QEC instructions.
Even small delays ($\sim 100ns$) in the implementation of quantum error correction can lead to a build up of errors that renders any computation useless.
Therefore alternative solutions are necessary. 

\begin{figure}[ht]
\includegraphics[scale=0.17]{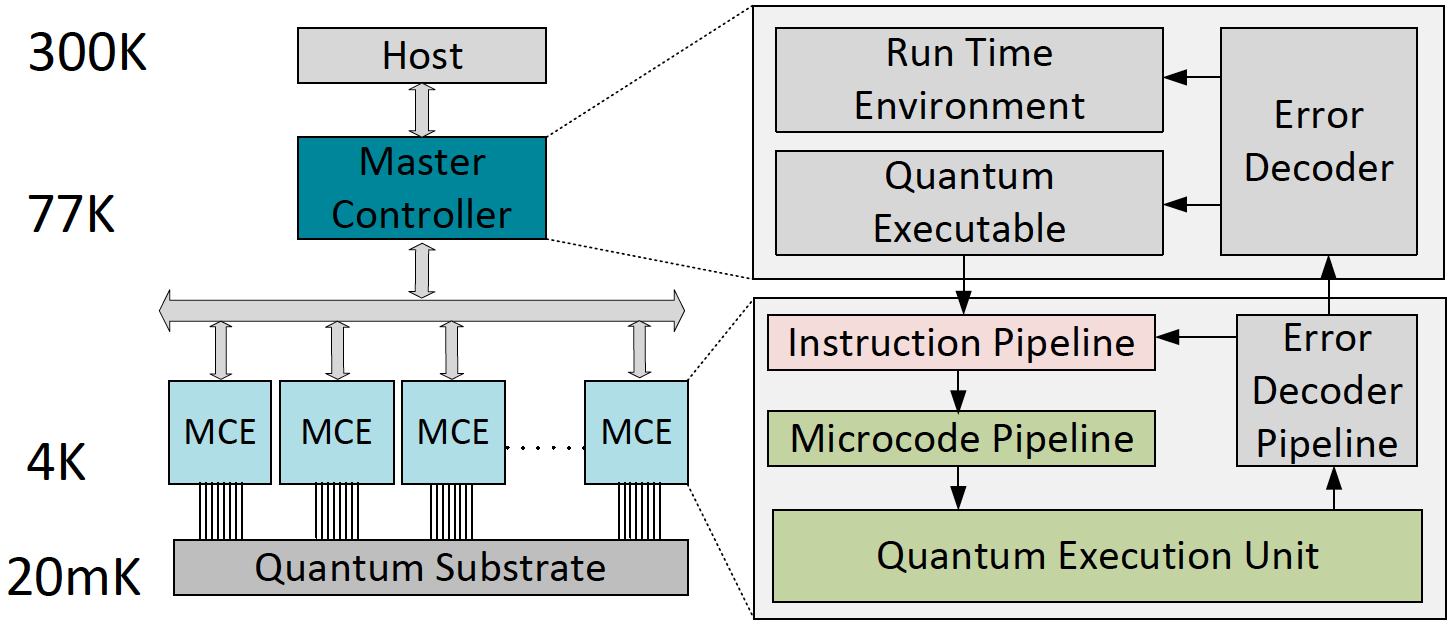}
\caption{The control processor consisting of the master controller and an array of MCEs, connected by a bus. Figure taken from \cite{tannu}.}\label{fig:stack}
\end{figure}

To overcome this instruction bandwidth bottleneck, an architecture that delegates the task of QEC from the software to the hardware has been proposed \cite{tannu}.
The control processor in this architecture consists of a master controller and an array of dedicated \emph{Micro-coded Control Engines} (MCEs), which are connected by
a global data and instruction bus (see Figure \ref{fig:stack}).
This architecture is designed to distribute the instruction delivery for QEC 
across the MCEs. Each MCE manages a dedicated ``tile" of a small number of qubits and executes the quantum error correction instructions without any
software coordination. 

An MCE consists of an instruction pipeline, a microcode pipeline, a quantum execution unit, and an error decoder pipeline. 
The instruction pipeline delivers logical instructions and translates them into physical instructions. 
The microcode pipeline feeds these instructions to the quantum execution unit. The microcode pipeline also stores the QEC instructions in memory and feeds these to the quantum execution unit as well.
The quantum execution unit executes the instructions that it receives. The error decoder pipeline is part of a two-stage decoding process. 
It collects the results of syndrome measurements and implements a simple look up table to correct single qubit errors.
More complicated errors are dealt with by a global decoder located in the master controller.

The microcode pipeline consists of a microcode memory, which stores the instructions for the QEC, and an address decoder. 
A micro-operation ($\mu\textnormal{op}$) corresponding to a QEC instruction moves from the microcode memory to the address decoder, which in turn 
delivers the $\mu\textnormal{op}$ to the quantum execution unit. Importantly, the capacity of the microcode memory affects the number of qubits that an MCE can manage. 
The necessary memory capacity of the MCE can be reduced by, in part, taking advantage of the repetitive nature of the instructions necessary
for the QEC \cite{tannu}. This increases the number of qubits managed by a single MCE by a factor of $90$. 

\begin{figure}[ht]
\includegraphics[scale=0.17]{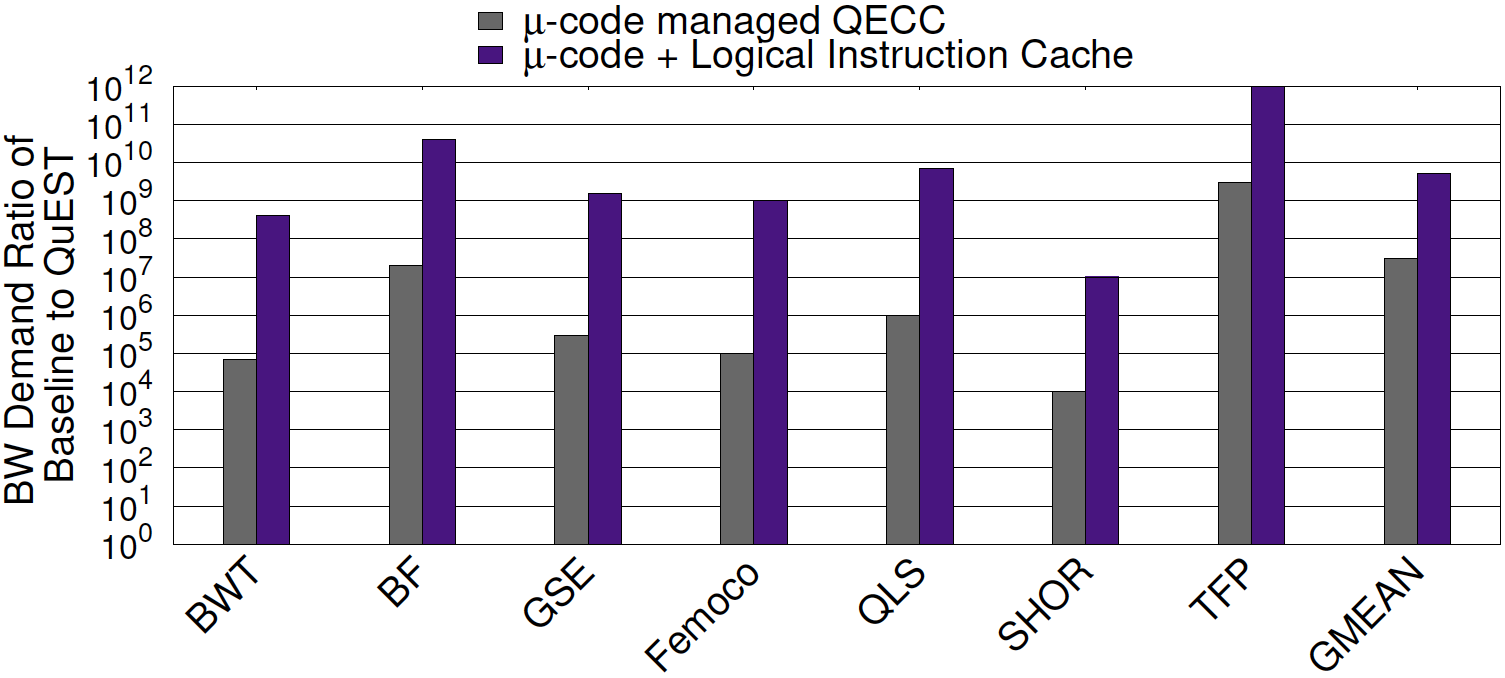}
\caption{Global bandwidth savings using local computation, taken from \cite{tannu}.}\label{fig:totalband}
\end{figure}

A comparison between the two architectures is given in \cite{tannu}, where QuRE is used to calculate the global instruction bandwidth requirements for seven quantum algorithms on each architecture. 
The baseline calculation represents the QEC being managed by software. In this case the compiler (or programmer) generates the physical instruction stream for the QEC as well as the logical instructions.
The architecture with dedicated MCEs to manage the instruction stream for the QEC is also costed, whose global bandwidth is 
comprised of the algorithms logical instructions as well as the master controller's synchronisation tokens.
Across the algorithms considered, a global bandwidth reduction of at least four orders of magnitude is observed (see Figure \ref{fig:totalband}).
One can also allow for the MCEs to cache the logical instructions; software managed instruction caches facilitate this process. This gives extra savings of three times an order of magnitude, which are also given in Figure \ref{fig:totalband}.

It is worth observing that the architectures were costed using several different syndrome extraction methods and different technologies. The values given in Figure \ref{fig:totalband} are for Steane style extraction using the projected gate latencies proposed by DiVincenzo \cite{DiVin}, which are often used in the QEC literature.
The other configurations of technology and extraction method produce very similar numbers (the coefficient of variance between the different configurations is 0.0002\%).

\section{Providing Access} \label{sec:conclude}

In this paper we have considered three bottlenecks that will have a substantial impact on near term quantum computing performance due to the divide between the QPU and CPU. In all three cases the link between the QPU and the CPU causes limitations i.e. latency introduced by the link or bandwidth restrictions. A possible way to mitigate these bottlenecks is to decrease the load on the link between the CPU and QPU by removing the restriction that classical computation has to be carried out on the CPU. We have shown that by moving some of the light-weight, regularly executed classical computation to the local computation unit within the QPU, these bottlenecks can be eliminated.

Moving away from carrying out classical computation solely on the CPU will require access to local computation within the QPU for algorithm and software developers. This will require a substantially different model for the execution of hybrid programs. In many near term quantum computers the local computation will be provided by FPGAs rather than the more traditional CPU. This is due to the strict timing requirements for operating the qubit I/O interfaces, for example, the triggering of a given pulse sequence. If we want to use local computation to mitigate bottleneck issues and to improve qubit utilisation, allowing developers to deploy gateware to these local computation units is vital albeit challenging. A number of proposals exist to provide the required access, see \cite{df, qm}.



\bibliographystyle{hunsrt.bst}
\bibliography{refs1}

\end{document}